\documentclass[structabstract]{aa}  

\usepackage{graphicx}
\usepackage{txfonts}
\usepackage{lscape}
\usepackage{natbib}
\usepackage{rotating}

\def\ie{i.e.}
\def\eg{e.g.}
\def\deg{\ifmmode^\circ\else$^\circ$\fi}

\begin{document}

\title{A minor merger origin for stellar inner discs and rings in spiral galaxies}

\author{M.~Carmen Eliche-Moral\inst{1}, A.~C.~Gonz\'{a}lez-Garc\'{\i}a\inst{2,3,4}, M.~Balcells\inst{3,4,5}, J.~A.~L.~Aguerri\inst{3,4}, J.~Gallego\inst{1}, J.~Zamorano\inst{1}, \& Mercedes Prieto\inst{3,4}}

\institute{Departamento de Astrof\'{\i}sica, Universidad Complutense de Madrid, E-28040 Madrid, Spain,  \email{mceliche@fis.ucm.es}
  \and
Dept.~F\'{\i}sica Te\'{o}rica, Universidad Aut\'{o}noma de Madrid, E-28049, Madrid, Spain
  \and 
Instituto de Astrof\'{\i}sica de Canarias, C/ V\'{\i}a L\'actea, E-38200 La Laguna, Tenerife, Spain
  \and
Departamento de Astrof\'{\i}sica, Universidad de La Laguna, E-38200 La Laguna, Tenerife, Spain
  \and
Isaac Newton Group of Telescopes, Apartado 321, E-38700 Santa Cruz de La Palma, Canary Islands, Spain
}

   \date{Received 14 January, 2011; accepted 19 May, 2011}

\abstract{
Recent observations show that inner discs and rings (IDs and IRs, henceforth) are not preferably found in barred galaxies, a fact that points to the relevance of formation mechanisms different to the traditional bar-origin scenario. In contrast, the role of minor mergers in the formation of these inner components (ICs), while often invoked, is still poorly understood.
}{
We have investigated the capability of minor mergers to trigger the formation of IDs and IRs in spiral galaxies through collisionless N-body simulations. 
}{
We have run a battery of minor mergers in which both primary and secondary are modelled as disc-bulge-halo galaxies with realistic density ratios. Different orbits and mass ratios have been considered, as well as two different models for the primary galaxy (a Sab or Sc). A detailed analysis of the morphology, structure, and kinematics of the ICs resulting from the minor merger has been carried out.
}{
All the simulated minor mergers develop thin ICs out of satellite material, supported by rotation. A wide morphological zoo of ICs has been obtained (including IDs, IRs, pseudo-rings, nested IDs, spiral patterns, and combinations of them), but all with structural and kinematical properties similar to observations. The sizes of the resulting ICs are comparable to those obtained in real cases with the adequate scaling. The existence of the resulting ICs can be deduced through the features that they imprint in the isophotal profiles and kinemetric maps of the final remnant, as in many real galaxies. Weak transitory oval distortions appear in the remnant center of many cases, but none of them develops a noticeable bar. The realistic density ratios used in the present models make the satellites to experience more efficient orbital circularization and disruption than in previous studies. Combined with the disc resonances induced by the encounter, these processes give place to highly aligned co- and counter-rotating ICs in the remnant centre. 
}{
Minor mergers are an efficient mechanism to form rotationally-supported stellar ICs in spiral galaxies, neither requiring strong dissipation nor the development of noticeable bars. The present models indicate that minor mergers can account for the existence of pure-stellar old ICs in unbarred galaxies, and suggest that their role must have been crucial in the formation of ICs and much more complex than just bar triggering.
}


\titlerunning{A minor merger origin for stellar inner discs and rings in spiral galaxies}
\authorrunning{Eliche-Moral et al.}

   \maketitle

\section{Introduction}
\label{sec:introduction}

Stellar IDs and IRs are found in at least one third of spiral galaxies \citep[][F06 hereafter]{2002AJ....124...65E,2006MNRAS.369..529F}. Usually masked by the light coming from brighter overlapping components, many of them have been detected through the peculiar features that these ICs imprint in the isophotes or in the kinematic maps of the host galaxy \citep[][]{1994AJ....108.1579V,1994ESOC...49..147K,1995AJ....110.2622L,1998A&AS..131..265S,1998MNRAS.300..469S,2000A&AS..145...71K,2001AJ....121.2431R,2004A&A...415..941E,2004hst..prop10157C,2006MNRAS.370..477L,2008A&A...478..403C,2010arXiv1004.2190M,2010AIPC.1240..251M,2010AIPC.1240..255S}. 

Stellar IDs are found in galaxies of all types, although they tend to reside in Sa-Sb's. Their radii span from some tens of parsecs to $\sim 2$\,kpc in diameter. They exhibit red or blue colors, usually similar to those of the host bulges, and harbour stellar populations with ages ranging from $\sim 1$\,Gyr to $\sim 10$\,Gyr \citep[][]{1988ApJ...327L..55F,1990A&A...229..441B,1993ApJ...407..525V,1994MNRAS.270..325C,2001A&A...367..405P,2002AJ....123..159C,2002AJ....124...65E,2002NewAR..46..187M,2003ApJS..146..299E,2004ARA&A..42..603K,2004MNRAS.354..753M,2005MNRAS.358.1477A,2007MNRAS.379..418C,2007MNRAS.379..401E,2007MNRAS.379..445P}.  Inner star-forming IRs and pseudo-rings are detected in at least one fifth of all disc galaxies \citep{2005A&A...429..141K,2009NewAR..53..169M,2011arXiv1103.5735D}. However, accounting for the episodic nature of their star formation histories and their long-lived stable configurations, the total fraction of IRs including pure stellar (non star-forming) ones is expected to be much higher \citep{2006ApJ...649L..79M,2007MNRAS.380..949S}. 

These stellar ICs could be primordial features in the galaxy centres, established at an early epoch of rapid, cold accretion flows at high redshift \citep{2006MNRAS.373.1125B}. However, the dramatic decline of the star formation surface density in the discs during the last $\sim 8$\,Gyr and the correlations found between the nuclear and disc global properties of spiral galaxies evidence an intertwined star formation history for bulges, discs, and ICs that spreads through time \citep{1996AJ....111.2238P,2008A&A...488.1167D,2008A&A...489.1003D}. Attending to the high detection fraction of stellar IDs and IRs in barred galaxies, their origin has been traditionally associated to bar patterns in discs  \citep{1992A&A...259L..27C,1995AJ....110..199A,1998MNRAS.298..267V,2002AJ....124...65E,2003ApJ...582..723R,2008AJ....135..479B,2008ApJ...684L..83D,2010MNRAS.402.2462C}. This has been corroborated by numerical simulations, that have shown that the dynamical resonances induced by bars give place to the rings surrounding them easily. Moreover, the bars induce strong gas inflows to the galaxy centre, which settle in circular orbits, inducing star formation that can result in an ID \citep{1996ApJ...462..114N,2009MNRAS.394...67A,2009MNRAS.400.1706A,2010MNRAS.402.2462C}. Other bar-related scenarios propose that IDs can also be the relics of diluted nested bars or of episodes of starbursts in rings that have shrunk in radius to the galaxy centre over time \citep{1992A&A...259L..27C,2003ApJ...582..723R,2010AstL...36..319S}. 

Nevertheless, recent observations indicate that stellar IDs and IRs are not preferably harboured by barred galaxies, being at least as frequent in non-barred early-type hosts as in barred ones \citep{2004MNRAS.352..721E,2004MNRAS.350...35F,2005A&A...429..141K,2006A&A...448..489K,2006MNRAS.366.1151S,2010MNRAS.402.2462C}. IRs are easily developed by unstable gas-rich discs in numerical simulations \citep{2010arXiv1007.0169A}, but IDs are structures more difficult to reproduce spontaneously  \citep{1997ApJ...479..702T}. So, other mechanisms capable of triggering disc resonances besides bars are required to explain the existence of stellar IDs and IRs in unbarred galaxies. One of the main alternatives is gas infall, but simulations indicate that this mechanism is efficient producing IRs, but not IDs  \citep{1998ApJ...506...93T,2006ApJ...636L..81N}. 

Another possible driver is merging, as suggested by the existence of many ICs exhibiting strong misalignments or distorted morphologies and/or kinematics with respect to the host disc \citep[see, \eg,][]{1994ASPC...59..376O,1995P&SS...43.1377A,1996FCPh...17...95B,1996ApJ...471..115B,2004A&A...423..481K,2005A&A...431..503R,2006ApJ...649L..79M,2006AJ....131.1336S,2009A&A...504..389C,2009A&A...507.1303F,2009NewAR..53..169M,2010MNRAS.401.2067B}. This scenario is supported by numerical simulations, which have proven that major mergers can drive the formation of kinematically-decoupled ICs analogous to the ones found in E-S0 galaxies  \citep[][]{1991Natur.354..210H,1998ApJ...505L.109B,2000MNRAS.316..315B,2001ASPC..240..135B,2007MNRAS.376..997J,2008A&A...477..437D}. Considering that massive E-S0 galaxies have experienced at least one major merger since $z\sim 1$ \citep{2009A&A...497...35G,2009ApJ...694..643L,2010A&A...519A..55E,2010arXiv1003.0686E}, this means that the existence of ICs in E-S0's can be satisfactorily explained through this process. However, they can not account for the ICs found in unbarred Sa-Sb galaxies, as the remnants resulting from major mergers are E-S0's for typical gas amounts \citep[][]{1996ApJ...471..115B,2004A&A...418L..27B,2005A&A...437...69B,2003ApJ...597..893N}. 

The straightforward alternative to major mergers is minor merging. Minor mergers induce dynamical resonances and oval distortions in discs easily and imprint a smooth growth to the pre-existing galaxy bulge \citep[see][]{1999ApJ...519L.127B,2002ApJ...573..131P,2006A&A...457...91E,2006AJ....131.1336S,2010MNRAS.401.2067B}. Although the effects of minor mergers on bulge and disc growth and on the satellites have been extensively studied  \citep[\eg,][]{2010ApJ...715..202H,2010MNRAS.403..768H,2010IAUS..262..432T,2010AIPC.1240..423T,2011arXiv1103.2562B,2011arXiv1103.2565E}, little attention has been devoted to study specifically their ability to induce the formation of dynamically-cold ICs in galaxies. 

One of the earlier studies dealing with this topic was carried out by \citet{1992A&A...257...17E}, who performed numerical simulations of ring-companion interactions to analyse the effects of the minor merger onto pre-existing outer rings. Later, Thakar and collaborators tested the formation of counter-rotating discs in spiral galaxies through retrograde mergers of gas-rich dwarfs onto discs \citep{1996ApJ...461...55T,1997ApJ...479..702T,1998ApJ...506...93T}. They found that, although counter-rotating thin gaseous discs were formed during the minor merger, the obtained sizes were only comparable to the biggest observational cases. The formed IDs did not have exponential profiles either and were highly unstable, quickly deriving into IRs. This led these authors to conclude that, in order to form a normal counter-rotating disc, "there must be either little or no pre-existing prograde gas in the primary galaxy, or its dissipative influence must be offset by significant star formation activity". More recently, \citet{2001A&A...367..428A} performed collisionless N-body simulations to test the growth of bulges after the accretion of dense spheroidal satellites. Their undisrupted dense satellite cores sank to the galaxy centre, giving place to kinematically-decoupled components in the remnants, but not supported by rotation.

\citet[EM06 hereafter]{2006A&A...457...91E} studied the effects of the accretion onto disc galaxies using satellites which themselves comprised a disc, a bulge, and a dark halo. More importantly, the relative densities of primary and secondary were made to be realistic by imposing that both models lie on the Tully-Fisher relation. This was an improvement over previous studies of mass buildup via accretion, given that the disruption of the satellite, the radius of deposition of satellite material, and the dynamical heating of the primary, all depend on the tidal fields, which scale with the relative densities of the two galaxies. EM06 reported the formation of dynamically-cold stellar structures in the centre of remnants, made out of disrupted satellite material. However, that paper was centred on the bulge growth driven by the minor merger, and hence no analysis or description of the ICs was performed. We have therefore extended the simulations of EM06, sampling different orbits and initial conditions and using $\sim 3$ times more particles, to carry out an exhaustive study of the morphology and kinematics of the ICs resulting from different minor mergers. We have simulated collisionless cases, as recent studies have demonstrated that dissipative components are not decisive in the formation and shaping of kinematically-decoupled ICs through major mergers \citep[although gas makes them more axisymmetric and can induce recent star formation in them, see][]{2007MNRAS.376..997J,2008A&A...477..437D}.

The present models demonstrate the capability of minor mergers to induce the formation of a wide morphological zoo of thin stellar ICs in the galaxy centres, with structural and kinematical properties analogous to those harboured by real spiral galaxies. The novelty of these models is two-fold: first, all the resulting ICs are made out of disrupted satellite material (whereas they come from resonances in the parent disc in previous studies) and, secondly, dissipative effects and strong bars are not essential to form these dynamically-cold ICs.

The paper is structured as follows. We briefly describe the models in \S\ref{sec:models}. The formed ICs are analysed geometrically, photometrically, and kinematically in \S\ref{sec:results}. In \S\ref{sec:observations}, we perform a qualitative comparison of the ICs resulting in the models with those detected in real spiral galaxies. Model limitations are commented in \S\ref{sec:limitations}. The discussion can be found in \S\ref{sec:discussion}. A brief summary of the results and some conclusions are finally addressed in \S\ref{sec:conclusions}. The physical magnitudes are provided in units of the simulation throughout the paper, although scaling to real galaxies is straightforward assuming the scalings described in \S\ref{sec:models}.

\section{The models}
\label{sec:models}

\begin{table}
\caption{Number of particles used in the models}
\label{tab:models1}
\centering
\begin{tabular}{lccccccc}
\hline\hline
 & \multicolumn{7}{c}{Number of Particles ($/10^3$)}\\\hline\vspace{-0.3cm}\\
 Experiment type & Total & D1 & B1 & H1 &   D2 & B2 & H2  \\
  \multicolumn{1}{c}{(1)}& (2)  & (3)    & (4)  & (5)   & (6)    & (7) &(8) \\\hline\vspace{-0.3cm}\\
Big bulge  & 185 & 40 & 10 & 90 &  10 & 5 & 30  \\
Small bulge  & 415 & 60 & 10 & 300 & ---  & --- & ---  \\\hline
\end{tabular}
\begin{minipage}[t]{0.48\textwidth}\vspace{0.1cm}
\emph{Columns}: (1) Experiment type depending on the used primary galaxy model (big or small bulge). (2) Total particle number. (3) Number of primary disc particles.  (4) Number of primary bulge particles. (5) Number of primary halo particles. (6) Number of satellite disc particles. (7) Number of satellite bulge particles. (8) Number of satellite halo particles. 
\end{minipage}
\end{table}

We have extended the set of collisionless N-body simulations of minor mergers onto disc galaxies described in EM06, now using longer pericenters and different initial disc galaxies. The outcome of the satellite material of a final set of 12 collisionless models has been analysed (six of them come from EM06). Ten of these experiments were run using a disc galaxy with a prominent bulge as primary galaxy (equivalent to a Sa-Sb galaxy), while in another two a primary with a smaller bulge (similar to a Sc) was used to investigate the influence of the primary bulge in the outcome of the accretion. 

All the galaxies in the simulations (primary galaxies and satellites) have an initial bulge-disc-halo structure. The primary galaxy models were built using the {\tt GalactICS} code \citep{1995MNRAS.277.1341K}, including an exponential disc component \citep{1969ApJ...158..505S}, a King bulge \citep{1966AJ.....71...64K}, and a dark halo built following an Evans profile \citep{1994MNRAS.269...13K}. The discs of both primary galaxy models follow an exponential surface density profile both radially and vertically, and were allowed to relax in isolation for about 10 disc dynamical times prior to placing them in orbit for the merger simulations. No relevant resonant structures appear in the disc. The primary galaxy with a big bulge matches the Milky Way (MW) when the units of length, velocity, and mass are $R=4.5$\,kpc, $v=220$\,km s$^{-1}$, and $M=5.1\times 10^{10} M_\odot$, respectively. In this case, the corresponding time unit is 20.5\,Myr. The primary galaxy with a small bulge matches NGC 253 using the following units of length, velocity, and mass: $R=6.8$\,kpc, $v=510$\,km s$^{-1}$, and $M=2.6\times 10^{11} M_\odot$, implying a time unit of 11.7\,Myr. These values, especially when using an appropriate $M_\mathrm{lum}/L$ ratio, yield mass-to-light ratio values close to observations ($M/L\sim 10$). Tables\,\ref{tab:models1} and \ref{tab:models1a} summarize the main characteristics of the two relaxed galaxy models used to represent the primary galaxy. For the remainder of the article we use model units in all the figures, a conversion to physical units is easily achieved through the above correspondences.

\begin{table*}
\begin{minipage}[t]{\textwidth}
\caption{Initial parameters of the primary galaxy models}
\label{tab:models1a}
\centering
\begin{tabular}{lccccccc}
\hline\hline
Experiment type & $\mathcal{M}_{\mathrm{T,1}}$ & $\mathcal{M}_{\mathrm{B,1}}/\mathcal{M}_{\mathrm{D,1}}$& $\mathcal{M}_{\mathrm{Dark,1}}/\mathcal{M}_{\mathrm{L,1}}$ & $h_{\mathrm{D,1}}$ & $r_{\mathrm{B,1}}/h_{\mathrm{D,1}}$ & $h_\mathrm{95\mathrm{\%,1}}/h_{\mathrm{D,1}}$ & $z_{\mathrm{D,1}}/ h_{\mathrm{D,1}}$ \\
  (1)& (2)  & (3)    & (4)  & (5)   & (6)    & (7) &  (8)\\\hline\vspace{-0.3cm}\\
Big primary bulge  &  6.40 & 0.51 &  4.16 & 1.00 & 0.20 & 3.7 & 0.11   \\
Small primary bulge  & 0.99 & 0.08 &  6.98 &  0.39   & 0.16  & 3.8 & 0.06 \\\hline
\end{tabular}
\begin{minipage}[t]{\textwidth}
\emph{Columns}: (1) Experiment type depending on the primary galaxy model used (big or small bulge). (2) Total primary mass (simulation units). (3) Bulge-to-disc mass ratio of primary galaxy. (4) Dark-to-luminous mass ratio of primary galaxy. (5) Radial disc scale-length of the primary galaxy. (6) Ratio between the effective radius of primary bulge and primary disc scale-length. (7) Ratio between the radius of the shell containing 95\% of luminous material in the primary galaxy and its disc scale-length. (8) Ratio between the vertical and radial disc scale-lengths of the primary galaxy.
\end{minipage}\end{minipage}
\end{table*}

Satellites are scaled replicas of the primary galaxy model with a big bulge in all the experiments. A physically-motivated size-mass scaling was used to ensure that the primary-to-satellite density ratios are realistic, forcing both galaxies to obey the Tully-Fisher relation \citep[][consult EM06]{1977A&A....54..661T}.  In Table\,\ref{tab:models2}, we list the characteristic masses and sizes of the satellites for each merger experiment. The luminous mass ratios between the primary galaxy and the satellite considered have been 1:6, 1:9, and 1:18. In the models with big primary bulges, the primary and satellite are scaled replicas, so the luminous mass ratio of the encounter is equivalent to its total mass ratio. However, as commented above, the satellites in the models with small primary bulges are scaled versions of the big primary bulge model, in order to analyse the influence of the primary bulge structure in the outcome of the accretion. Satellites were also relaxed in isolation prior to the merging simulation.

The initial separation of both galaxies was 15 primary disc scale-lengths in all the experiments. In order to avoid a perfect spin-orbit coupling, the initial inclinations between the orbital plane and the galactic planes of the primary and the satellite galaxies were fixed to 30\deg\ in direct orbits and to 150\deg\ in retrograde orbits, respectively. All satellites have also an azimuthal angle $\phi=90\deg$. For each satellite mass, pericenter distance, and model of the primary galaxy considered, a direct and a retrograde orbit have been computed. Initial orbits were elliptical with pericenters equal to one or eight disc scale lengths, depending on whether we were running a short or a long pericenter orbit. Relative velocities at the first pericenter passage oscillate between 440--650 km/s, considering the scalings commented above. 

In Table\,\ref{tab:models2}, we include the orbital parameters of each merging experiment. As satellite models are scaled down versions of the primary model with a big bulge and exhibit different mass ratios having the same number of particles, this renders different mass particles for the different
components in each experiment. The largest mass contrast, and thus the higher two-body errors, are expected in the models with a total mass ratio of 1:18, where a primary halo particle is 10 times more massive than the bulge particles of the satellite. Such extreme ratio is well below the limits explored earlier in different simulations using a similar set of initial models, so the kinematics of the inner regions of the remnant are much better sampled in the present models than
previously \citep[][EM06]{1998ApJ...505L.109B,2005MNRAS.357..753G}. The difference in the number of particles for the haloes hosting a big and a small bulge is two folded. On one hand, a bulge-less disc model stable to bar distortions requires a concentrated halo \citep[see, e.g.,][]{2005MNRAS.357..753G}. On the other hand, the masses of the halo and bulge particles need to be of the same order for ensuring low two-body errors. These two constraints required the models with small bulge to have such a high number of halo particles. As we are interested in the physics of the inner regions of the remnants, we intended to balance between the accuracy gained by using a higher number of particles and the economy of
resources in building new stable models for each satellite.

The evolution of the new models was computed using GADGET-2 code \citep{2001NewA....6...79S,2005MNRAS.364.1105S}. We used a softening of $\varepsilon=0.02$ in model units and an opening angle of $\theta = 0.6$. Considering this tolerance parameter and applying quadrupole-moment corrections, the code computes forces within 1\% of those given by a direct summation and preserves total energy better than 0.1\%. We evolved all models for $\sim$4 halo crossing times beyond full merger to allow the final remnants to reach a quasi-equilibrium state with a good conservation of energy in all runs. Times to full merger and total run times of each experiment are also indicated in Table\,\ref{tab:models2} in simulation units. The initial $B/D$ ratios of the main galaxy increase after the merger to $B/D\sim 0.6$ in the cases of big bulges and $B/D\sim 0.2$ for small bulges. This means that the remnants of the experiments starting with a Sb primary galaxy have become S0-Sa's, while the Sc primary galaxies have been transformed into Sb-Sbc's after the minor merger \citep[see][]{2001AJ....121..820G}.  

We will refer to each model throughout the paper according to the following code: M$m$P[l/s][D/R][b/s], where $m$ indicates the bulge-to-satellite mass ratio ($m=6$, 9, or 18 for models with luminous mass ratios equal to 1:6, 1:9, 1:18, respectively), "Pl" indicates long pericenter and "Ps" short pericenter, "D" or "R" describes the orbit ("D" for direct and "R" for retrograde), and the final "b" or "s" letter indicates if the primary galaxy had a big or a small bulge (see Table\,\ref{tab:models2}). 

\begin{table*}
\begin{minipage}[t]{\textwidth}
\caption{Orbital and scaling parameters of each merger experiment}
\label{tab:models2}
\centering
\begin{tabular}{llccccrcc}
\hline\hline
\multicolumn{1}{c}{Model Code} & \multicolumn{1}{c}{Code in EM06} & Primary Bulge& $\mathcal{M}_{\mathrm{L,Sat}}/\mathcal{M}_{\mathrm{L,Prim}}$  & $R_{\mathrm{Sat}}/R_{\mathrm{Prim}}$ & $R_{\mathrm{pericenter}}/h_\mathrm{D,1}$  &\multicolumn{1}{c}{$\theta _\mathrm{Prim}$} &  $t_\mathrm{full\,merger}$ & $t_\mathrm{total}$\\
\multicolumn{1}{c}{(1)}    & \multicolumn{1}{c}{(2)} & (3)         & (4)    & (5)        & (6)      & \multicolumn{1}{c}{(7)}      & (8)  & (9)\vspace{0.05cm}\\\hline\vspace{-0.3cm}\\
 (a) M6 Ps Db & M2TF35D & Big (b)  & 1:6 (M6) &  0.46  &  0.73 (Ps) &  30 (D)   & $\sim$72 & 100 \\
 (b) M6 Ps Rb & M2R     & Big (b)  & 1:6 (M6) &  0.46  &  0.73 (Ps) & 150 (R)  & $\sim$80 & 100 \\
 (c) M6 Pl Db & \multicolumn{1}{c}{---} & Big (b)  & 1:6 (M6) &  0.46  &  8.25 (Pl) &  30 (D)   & $\sim$93 & 144\\
 (d) M6 Pl Rb & \multicolumn{1}{c}{---} & Big (b)  & 1:6 (M6) &  0.46  &  8.25 (Pl) &  150 (R)   & $\sim$110& 144\\
 (e) M6 Ps Ds & \multicolumn{1}{c}{---} & Small (s)& 1:6 (M6) &  0.25  &  0.87 (Ps) &  30 (D)   & $\sim$40 & 62\\
 (f) M6 Ps Rs & \multicolumn{1}{c}{---} & Small (s)& 1:6 (M6) &  0.25  &  0.87 (Ps) & 150 (R)  & $\sim$44 & 72\vspace{0.05cm}\\\hline\vspace{-0.3cm}\\
 (g) M9 Ps Db & M3TF35D & Big (b)  & 1:9 (M9) &  0.39  &  0.79 (Ps) &  30 (D)   & $\sim$80 & 100 \\
 (h) M9 Ps Rb & M3R & Big (b)  & 1:9 (M9) &  0.39  &  0.79 (Ps) & 150 (R)  & $\sim$87 & 100 \vspace{0.05cm}\\\hline\vspace{-0.3cm}\\
 (i) M18 Ps Db & M6TF35D & Big (b)  & 1:18 (M18)&  0.28 &  0.86 (Ps) &  30 (D)   & $\sim$116& 122 \\
 (j) M18 Ps Rb & M6R & Big (b)  & 1:18 (M18)&  0.28 &  0.86 (Ps) & 150 (R)  & $\sim$142& 154\\
 (k) M18 Pl Db & \multicolumn{1}{c}{---} & Big (b)  & 1:18 (M18)&  0.28 &  8.19 (Pl) &  30 (D)   & $\sim$225& 260 \\
 (l) M18 Pl Rb & \multicolumn{1}{c}{---} & Big (b)  & 1:18 (M18)&  0.28 &  8.19 (Pl) &  150 (R)   & $\sim$285& 340 \\\hline\\
\end{tabular}
\begin{minipage}[t]{\textwidth}
\emph{Columns}: (1) Model code: M$m$P[l/s][D/R][b/s], where $m$ indicates the bulge-to-satellite mass ratio ($m=6$, 9, or 18 for models with luminous mass ratios equal to 1:6, 1:9, 1:18, respectively), "Pl" refers to long pericenter and "Ps" to short pericenter, "D" indicates direct orbits and "R" retrograde orbit), and the final letter ("b" or "s") indicates if the primary galaxy had a big or a small bulge. The letter in parentheses helps to identify each model quickly in the forthcoming figures. (2) Model code in EM06, for those models that were already presented in that paper. (3) Primary galaxy model used in the experiment (big or small primary bulge, see Table\,\ref{tab:models1}). (4) Luminous mass ratio between satellite and primary galaxy. (5) Ratio between the luminous half-mass radii of the satellite and the primary galaxy. (6) First pericenter distance of the orbit, in units of the primary disc scale-length. (7) Initial angle between the orbital momentum and the primary disc spin. This angle determines if the orbit is prograde (direct) or retrograde. (8) Approximate time of full merger, in simulation units. (9) Total run time of each experiment, in simulation units.
\end{minipage}
\end{minipage}
\end{table*}

\begin{figure*}[t]
\begin{center}
\includegraphics*[width=0.49\textwidth]{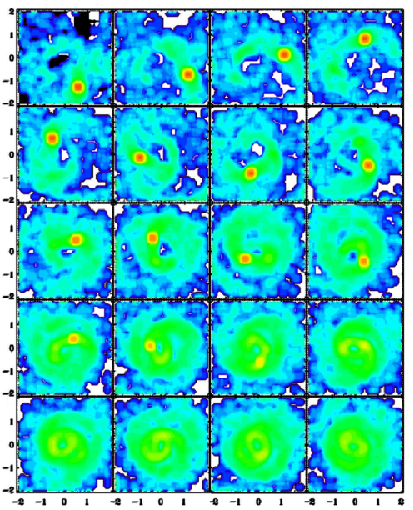}
\includegraphics*[width=0.49\textwidth]{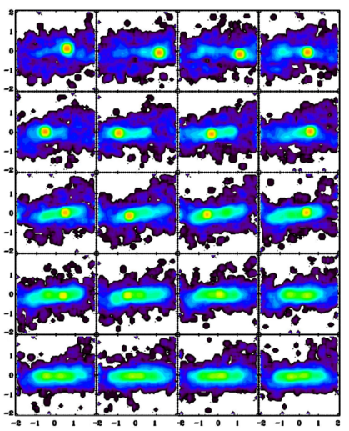}
\caption{Time evolution of the luminous surface density of the satellite material during the last moments of experiment M18PsDb (experiment "i" in Table\,\ref{tab:models2}). Face-on and edge-on views centred in the initial primary galaxy are plotted  (left and right panels, respectively). Snapshots corresponding to times from 101 to 120 are shown from up-to-down and left-to-right of each figure, using a time step equal to one. If the primary galaxy is scaled to the MW (see \S\ref{sec:models}), the total time period represented corresponds to $\sim 0.5$\,Gyr. A rainbow colour palette is used to represent different surface density levels in logarithmic scale, with redder colors indicating higher values. Spatial scales in both axes are provided in simulation units. The disrupted satellite material is finally deposited in the remnant centre forming an IR (in yellow) embeded into a more extended ID (in green). The IR rotates in the same direction as the primary disc material.}\label{fig:densevo1}
\end{center}
\end{figure*}

\section{Results}
\label{sec:results}

In EM06, we reported the formation of inner dynamically-cold components in minor merger experiments, but we did not analyse the resulting ICs neither structurally nor kinematically. In the next sections,  a detailed analysis of the co- and counter-rotating ICs formed in the models described in \S\ref{sec:models} is performed.

\subsection{Formation of ICs}
\label{sec:formation}

In Fig.\,\ref{fig:densevo1} we show the disruption experienced by the satellite in model M18PsDb (model i in Table\,\ref{tab:models2}) as an example of the time evolution of the luminous surface density of the satellite material during the last moments of the encounter. Although the different initial conditions affect to the global shape, size, and even the number of components in these inner structures, the final structure in all the experiments resembles a central vertically thin torus or disc (depending on whether the satellite material reaches the remnant centre or not), embeded into a more extended flat component, similar to a more extended disc. In the model plotted in the figure, the toroidal structure corresponds to the central ring visible at $R\sim 0.5$, while the outer disc corresponds to the low density structure that extends up to $R\sim 1$--1.5. The morphologies of the ICs formed in each experiment are described in detail in \S\ref{sec:zoo}.

Figure\,\ref{fig:densevo1} also shows that the outer and inner regions of the structure resulting from the satellite disruption are built up at different epochs during the minor merger. The less bound particles of the satellite (\ie, those from the disc) are disrupted earlier in the interaction. During the first pericenter cross of the orbit, its outer shells are removed from the satellite by the primary galaxy tidal field, giving place to the outer structure of the formed IC (as observed in the first half of snapshots in Fig.\,\ref{fig:densevo1}). The core of the satellite takes more time to experience a noticeable disruption. Its material is deposited at inner radii during the last stages of the encounter.

\subsection{Geometrical and kinematical characterization of ICs}
\label{sec:geometrical}

In this section, we describe the geometry and structure of these ICs through their radial and vertical surface brightness profiles, and analyse their misalignment with respect to the galaxy plane of the final remnant and their kinematics.

\begin{figure*}[!ht]
\begin{center}
\includegraphics*[width=0.85\textwidth]{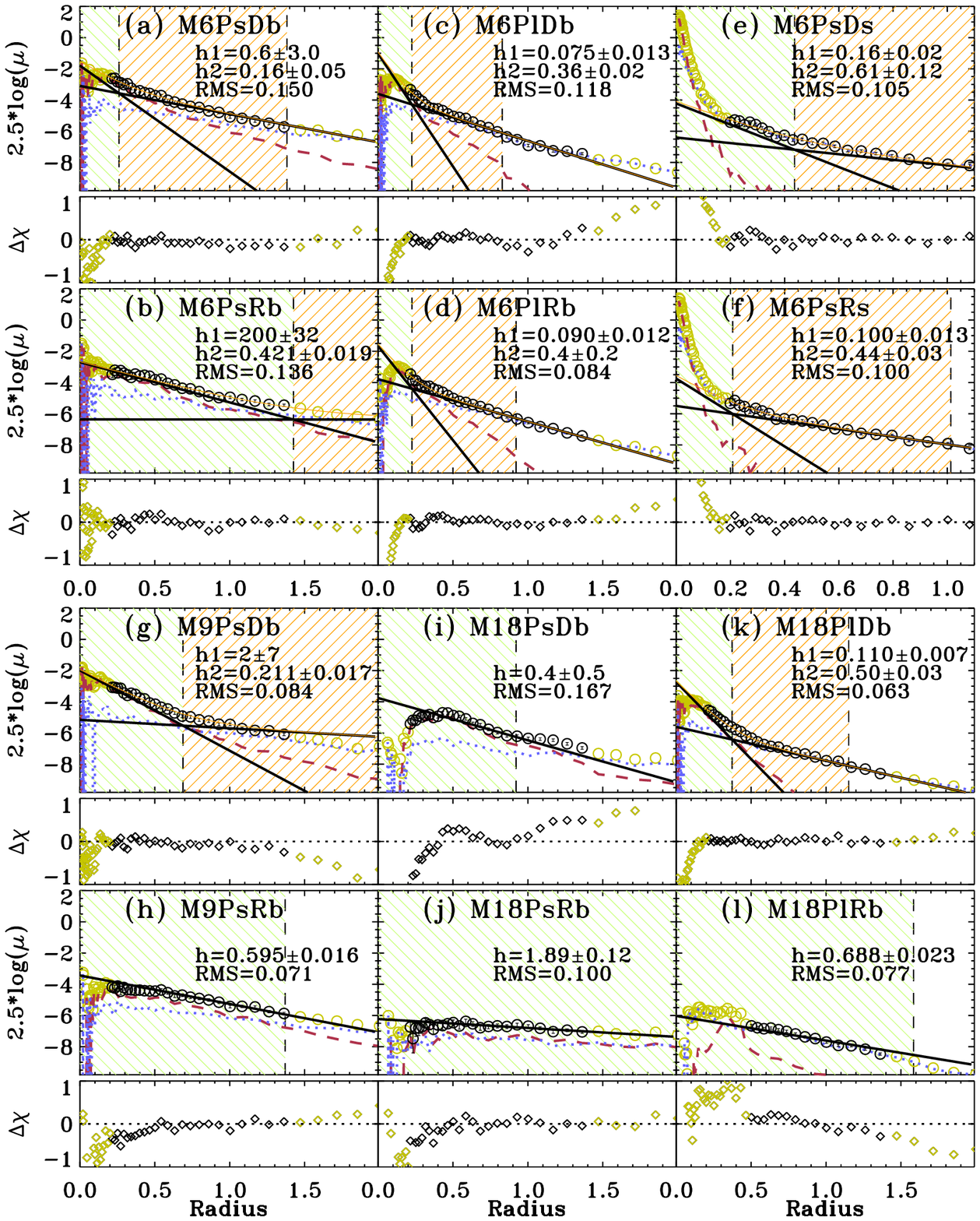}
\caption{Radial surface brightness profiles of disrupted satellite material in the final remnants of all the models. The plotted physical magnitudes are provided in simulation units. The letters identifying each panel are those used in Table\,\ref{tab:models2}. \emph{Circles}: Considering all the stars originally belonging to the satellite. \emph{Red dashed lines}: Satellite bulge stars. \emph{Blue dotted lines}: Satellite disc stars. \emph{Black solid straight lines}: Exponential fits to the radial density profiles considering all the satellite stars. The data points plotted in green have been excluded from the fits. In the panels corresponding to remnants that are better fitted with two exponential discs, the two scale-lengths are shown in the frame. When just one exponential disc provides a better fit to the global satellite stellar profile than two nested ones, only one scale-length is provided. The root mean square of the global fit is also shown in each remnant. \emph{Orange light solid curved lines}: Total fit to the radial surface density profiles in those cases that are better fitted by two exponential discs. \emph{Vertical dashed lines}: Extent of each IC characterized by each fitted exponential radial profile, as defined in the text. The radial extension of the ICs is remarked using a shaded background with different color for the different ICs identified in each model (green for the innermost ones and orange for the outer ones). \emph{Sub-panels}: Relative residuals of the fits as a function of the radial position. The models with small primary bulges obey a different scaling to the 
rest of models (see \S\ref{sec:models}). Therefore, the simulation units in these two experiments differ from those in the others, as observed in Figs.\,\ref{fig:sbrs2}-\,\ref{fig:mosaicmorph1}.}\label{fig:sbrs2}
\end{center}
\end{figure*}

\subsubsection{Radial distribution of ICs in the remnants}
\label{sec:sbr}

Figure\,\ref{fig:sbrs2} shows the radial surface density profiles of luminous material initially belonging to the satellite, to its disc, and to its bulge in the final remnants of all the experiments. In general, the inner structure is basically composed by satellite bulge particles (typically, at $R\lesssim h_\mathrm{D,1}(t=0)$), while the outer parts are controlled by the surface density of satellite disc particles. However, there is no sharp transition between the two radial ranges where each satellite component dominates. 

We have fitted one exponential disc or a combination of two nested ones to the surface density profiles of the satellite stellar material in each remnant (represented by two solid black straight lines in each panel). We assume that the satellite material described by each exponential radial profile defines an independent structure inside the global IC formed in the remnant center out of disrupted satellite material. In the fits, we have rejected the central regions of the density profiles to avoid the particular features exhibited by each distribution in the centre, such as undisrupted satellite cores (as in M6Ps[D/R]s, models e and f), holes (as in M18P[l/s]Db, models i and k), and smooth-length effects. The resulting fits and their residuals are plotted in the corresponding panels and sub-panels of Fig.\,\ref{fig:sbrs2}. 

Most of the resulting ICs can be well-approximated by one exponential radial profile or by the addition of two ones. However, notice that the independent ICs identified by each fitted exponential radial profile are shaped with material of both satellite components: from the bulge and the disc. Only in the experiments with a small primary bulge, the satellite bulge ends undisrupted in the remnant centre (see the profiles corresponding to the satellite bulge particles in M6PsDs and M6PsRs, frames e and f in the figure). In these models, the outer shells of satellite material configure an extended ID structure hosting the undisrupted satellite core. 

In order to delimit the radial extent of the ICs characterized by an unique exponential profile, we have assumed that, in the case that only one exponential profile is required to explain the whole radial structure, this IC extends up to the radius at which the fitted surface density is equal to 1/10 of its central value as extrapolated from the fit. In the case that the structure is better described by two nested exponential profiles, the innermost one is considered to extend up to the radius where the outer one starts to dominate the global fit. The outer one will extend from this radius up to that at which its surface density drops to 1/10 of its central value. The radial extent of each radial component resulting from these criteria are marked in Fig.\,\ref{fig:sbrs2}.

Summarizing, all the simulated minor mergers give place to complex extended ICs in the remnants made out of disrupted satellite material, with radial surface density profiles that can be well-described by one or two nested exponential profiles.

\subsubsection{Vertical distribution of ICs in the remnants}
\label{sec:hz}

An IC with a radial exponential profile can correspond to an ID or to a bulge with S\'{e}rsic index $n=1$ (i.e., a pseudo-bulge). The difference between all these components arise in the vertical thickness of the component, as compared to its own radial scale-length and to the one of the host disc. Following \citet{2011arXiv1103.1692S}, a ratio of scale-length to scale-height of about 3 is a reasonable frontier between spheroids and discs. So, we have estimated the ratio between the vertical and the radial scale-lengths of the formed ICs in each remnant, as well as the ratio of their scale-heights to the radial scale-length of the remnant discs, in order to identify them as IDs (thin or thick) or pseudo-bulges.

Figure\,\ref{fig:zscale} presents the characteristic vertical scale-lengths $h_\mathrm{z}$ at different radial positions of the disc, bulge, and all luminous material originally belonging to the satellite in the final remnants. The scale-lengths have been derived from exponential fits to the vertical density profiles at each radius. Only the radial positions with enough particles to ensure a vertical density profile of $S/N>50$ have been considered (typically, $R\lesssim 2\, h_\mathrm{D,Primary}$). The fitting errors are on average $\lesssim 10$\% for each estimate. The final scale-length of the disc remnant is marked in each panel as a reference (horizontal dotted lines). We have also indicated the radial extent of each IC, according to the definition adopted in \S\ref{sec:sbr}. 

Notice that the IC made out of satellite material presents scale-heights smaller than the radial scale-length of the original primary disc along the whole considered radial range in all the models, although it is vertically wider than the structure made out of primary disc material at all radii in all the remnants (compare asterisks and squares in Fig.\,\ref{fig:zscale}). So, although the primary disc has been heated by the satellite accretion, it is thin as compared to the IC. Therefore, the satellite material contributes to a structure thicker than the main remnant disc. This result is coherent with the structures of disc galaxies resulting from $\Lambda$CDM cosmological simulations, where the disrupted satellites contribute to the buildup of the thick disc of the galaxy \citep{2003ApJ...597...21A}. In general, the scale-height of the structure made of disrupted satellite material increases with radial position, meaning that these structures are flared (see asterisks in the figure).

The material of the disrupted satellite bulge tends to exhibit lower scale-heights than that of the disrupted satellite disc, but the different disruption epochs of both components seems to not affect to their final vertical distributions at a given radial position, as the scale-heights of both distributions are similar at each radius (compare red diamonds and blue triangles in the figure).

In Fig.\,\ref{fig:zscale}, we also show the ratio between the vertical and radial scale-lengths of each IC (numbers in black characters in each frame). Nearly half the innermost ICs formed in the models with big primary bulges are thin, as they exhibit ratios typically below $\sim 0.4$. This means that these  innermost components are IDs, according to \citet{2011arXiv1103.1692S} criterion. Other cases exceed the limiting value proposed by these authors (as models M6P[s/l]Rb, panels b and d in the figure), meaning that these ICs would be classified as pseudo-bulges in the case that the original primary galaxies lacked of a large central component. As this is not the case (they had a massive central bulge), the ICs formed in these models are embeded into the pre-existing bulge or the galaxy thick disc. We also have some questionable cases with vertical-to-radial scale-lengths near the limiting value for distinguishing between IDs and pseudobulges, as models M[9/18]PsRb (panels h and j in the figure).

The ratios of the scale-heights of these innermost components to the radial scale-length of the remnant disc indicate that the majority of these innermost ICs exhibit ratios below 0.4 (see the numbers in blue characters at the top left of each panel in the figure). This means that these structures are buried in the final remnant disc. The models with small primary bulges (panels e and f in the figure) exhibit high scale-heights in the center, because the satellite core sinks to the remnant center without experiencing disruption. None of these two models develops a pseudobulge out of disrupted satellite material, because the undisrupted satellite core produces a central peak in the radial surface density profile with a S\'{e}rsic index $n\sim 2$ (see the corresponding panels in Fig.\,\ref{fig:sbrs2}). Some of the innermost ICs also present a central hole in their radial surface density profiles (see panels d, i, and l in Fig.\,\ref{fig:sbrs2}). So, considering their low thickness, they must be IRs instead of IDs (this is confirmed by surface density maps in \S\ref{sec:zoo}).

The outer ICs are thick in general, following the \citeauthor{2011arXiv1103.1692S} criterion, so we can conclude that these structures basically shape the thick disc in the remnants.

Therefore, the minor mergers give place to ICs in the remnant centers out of satellite material that can be described as simple IDs, nested IDs, IRs hosted by IDs, and bulges hosted by IDs. Although their thickness depends on the case, the central regions of the ICs are generally embeded in the remnant disc and are relatively thin, while the outer ones constitute a thick flared discy structure surrounding the main remnant disc. The global structural characteristics of each IC as derived from their radial and vertical density profiles in Figs.\,\ref{fig:sbrs2}-\ref{fig:zscale} are summarized in Table\,\ref{tab:ics}.

\subsubsection{Morphology of ICs}
\label{sec:zoo}

\begin{figure}[!h]
\begin{center}
\includegraphics[width=0.48\textwidth]{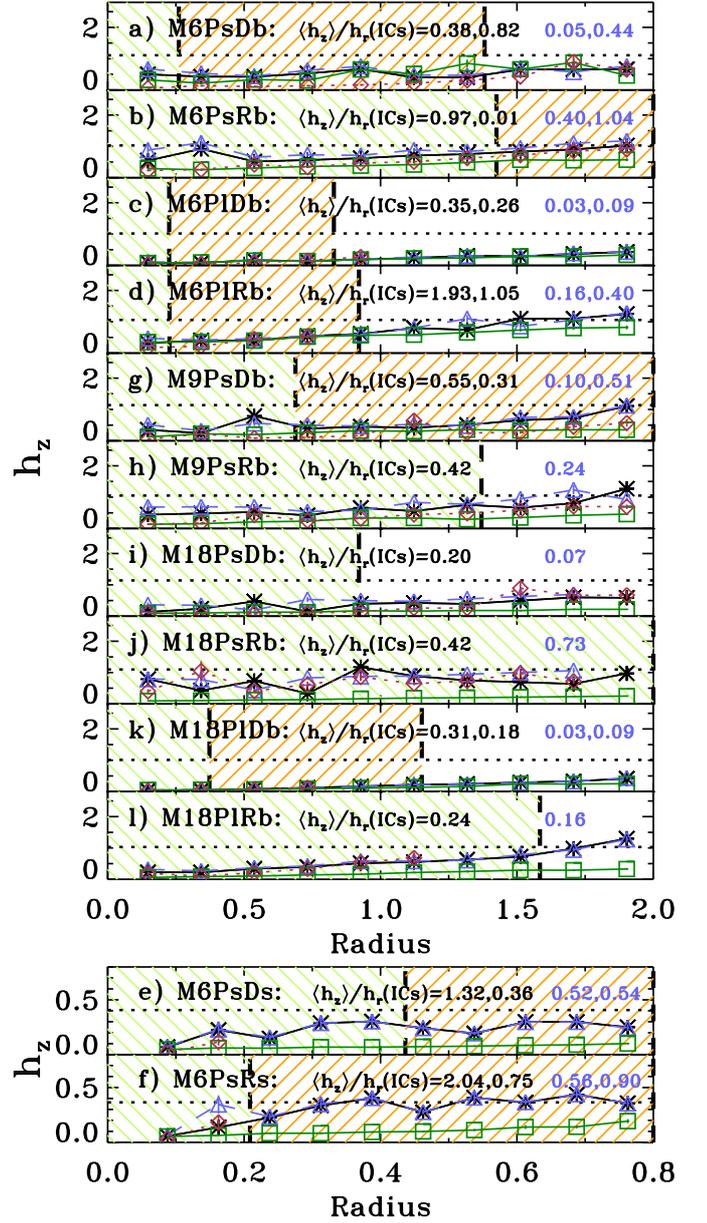}
\caption{Vertical exponential scale-lengths of material in the final remnants as a function of radius. The plotted physical magnitudes are provided in simulation units. The letters identifying each panel are those used in Table\,\ref{tab:models2} for each model. \emph{Triangles}: Material originally from the satellite disc. \emph{Diamonds}: From the satellite bulge. \emph{Asterisks}: Stellar material originally in the satellite. \emph{Squares}: From the original primary disc. \emph{Horizontal dotted line}: Scale-length of the final remnant disc, marked as a reference. \emph{Vertical dashed lines}: Radial extent of each IC characterized by an unique exponential profile, according to Fig.\,\ref{fig:sbrs2}. The ratio of the vertical to the radial scale-lengths of each IC is indicated in black characters (one number per radial exponential ICs). The ratio of the scale-height of each IC to the radial scale-length of the final remnant disc is also indicated in blue characters at the top left of each panel.}\label{fig:zscale}
\end{center}
\end{figure}

Figures\,\ref{fig:mosaicmorph1}-\ref{fig:mosaicmorph2} show the morphology of the ICs formed in each remnant. In order to provide a 3D description of the ICs structure, we plot the face-on and edge-on surface density maps of the disrupted stellar satellite material in all the remnants. The figure shows that the satellite disruption in our experiments has given place to a wide zoo of ICs, all made out of disrupted satellite material, and confirms the global morphology derived from the radial and vertical surface density profiles of Figs.\,\ref{fig:sbrs2}-\ref{fig:zscale} (see Table\,\ref{tab:ics}).

Moreover, the maps reveal the existence of substructures than make the morphology of the IC even more complex, such as ring relics (M18PlDb, panel k in Fig.\,\ref{fig:mosaicmorph2}), pseudo-rings (i.e., a non-closed ring-like structure like the one developed in experiment M18PlRb, panel l in Fig.\,\ref{fig:mosaicmorph2}), and spiral arms (M18PsDb, panel i in Fig.\,\ref{fig:mosaicmorph2}). The twin clumps observed in some ICs could be tracing weak bars or relics of oval distortions (see panels a, c, and d in Fig.\,\ref{fig:mosaicmorph1} and panels h-i in Fig.\,\ref{fig:mosaicmorph2}). The global ellipticity and PA profiles of the final remnants seem to support the bar-related origin of some of these structures (see \S\ref{sec:ellipticity}). However, none of the remnants has developed a clear nuclear bar in the IC, despite of having IRs and pseudo-rings. We summarize the substructures found in each case in Col.\,3 of Table\,\ref{tab:ics}. 

Our models demonstrate that minor mergers onto disc galaxies can give place to dynamically-cold thin ICs, with varied morphologies ranging from nuclear bars embeded in nested IDs to pseudo-rings, covering all different kinds of inner structures.

\subsubsection{Bars and oval distortions induced by the minor merger}
\label{sec:bars}

\begin{figure}[t]
\begin{center}
\includegraphics[width=0.5\textwidth]{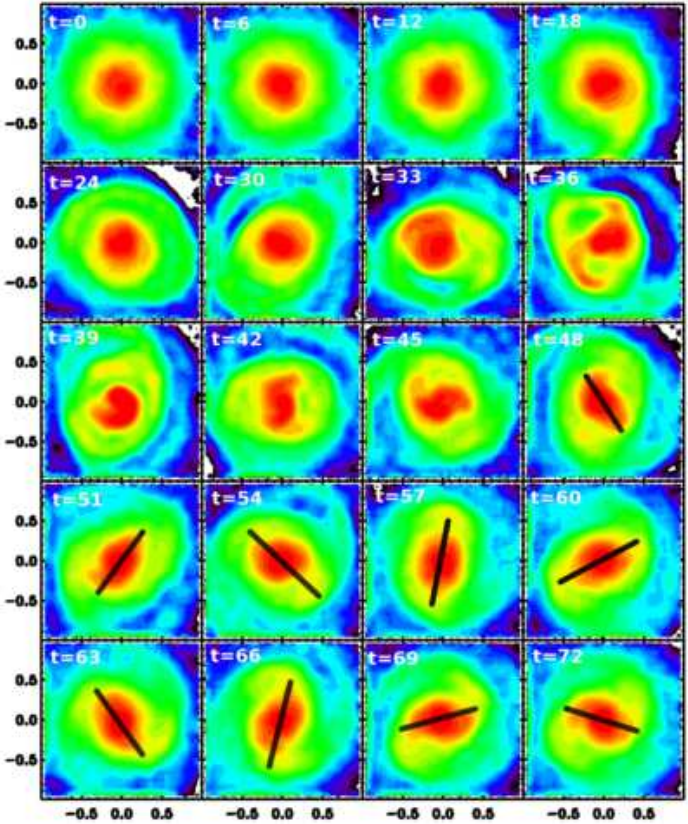}
\caption{Formation of a transitory bar-like distortion in the primary disc material during a minor merger. The time evolution of the surface density map of the primary disc material of model M6PsRs is plotted, using a face-on view of the initial primary galaxy (experiment f in Table\,\ref{tab:models2}). Time is shown at the top left corner of each panel in simulation units. A rainbow colour palette is used to represent different surface density levels in logarithmic scale, with redder colors indicating higher values. All physical quantities are given in simulation units. The major axis of the oval distortion formed in the remnant disc by the satellite impact is marked with a black line in the last panels. It rotates in the same direction as the primary disc material (clockwise). The orbit and rotation of the accreted satellite is counter-clockwise. The life time of this bar-like distortion corresponds to $\sim 0.3$\,Gyr, using the scaling indicated in \S\ref{sec:models}.}\label{fig:bar}
\end{center}
\end{figure}

Numerical simulations have shown that tidal interactions easily produce bars and non-axisymmetric distortions (such as ovals) in thin discs through gravitational torques  \citep[see][]{2005A&A...438..507B,2005MNRAS.364..607M,2009A&A...494..891A}. The time evolution of the density maps of the primary disc material reveals that our simulations are not exceptions: elongated distortions in privileged directions, defined by the orbit of impact, are induced in the primary disc by the minor merger (see Fig.\,\ref{fig:bar}). Some of these transient bar-like distortions remain until the end of the simulation, but extremely weakened (as in the case of the figure), but most of them dissolve during the last stages of the remnant relaxation. 

The strongest and longest-lived central oval distortions induced by the minor merger in the primary disc appear in the cases with a small primary bulge, corroborating many previous studies that have posed that a high mass concentration in the galaxy centre tends to stabilize the disc, preventing self gravity and thus bar formation \citep[see, \eg,][]{1990ApJ...363..391P,2002A&A...392...83B,2005MNRAS.363..496A,2006A&A...457...91E,2008MNRAS.384..386C}. All the transitory bar-like or oval patterns formed in the primary discs rotate in the same direction as the primary disc stars even in the retrograde mergers.

\begin{figure*}[t]
\begin{center}
\includegraphics*[width=\textwidth]{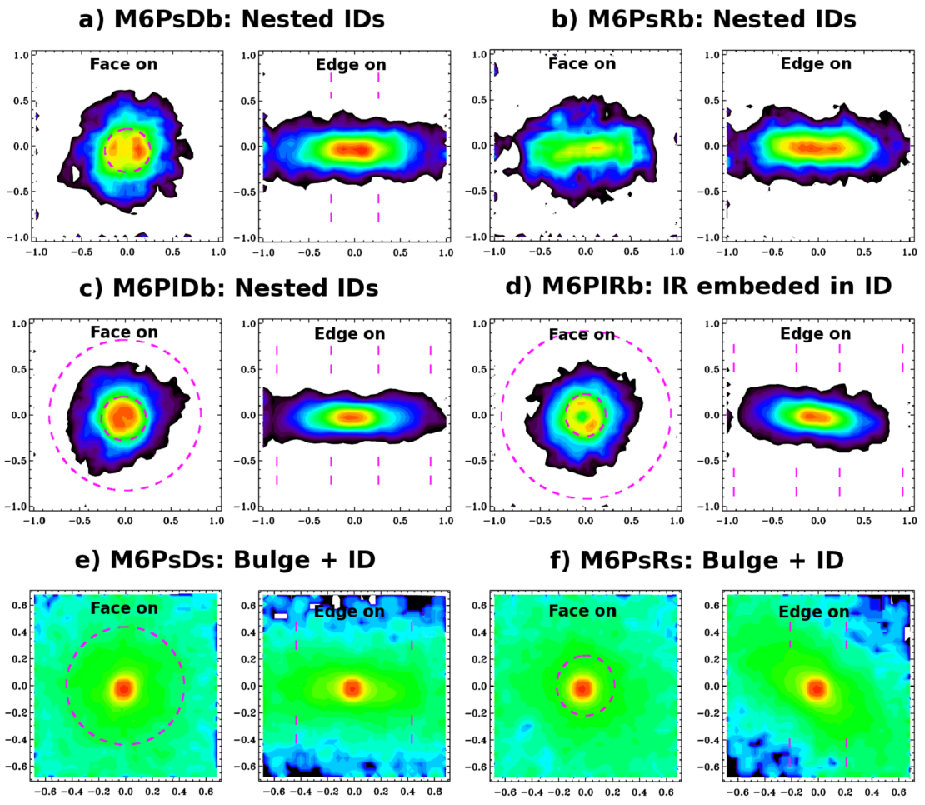}
\caption{Morphology of the ICs made out of disrupted satellite stellar material resulting in models with mass ratio 1:6 (models a-f in Table\,\ref{tab:models2}). Surface density maps of the stars originally from the satellite in the final remnants are presented, using a face-on and an edge-on view (left and right columns in each model, respectively). A rainbow colour palette is used to represent different surface density levels in logarithmic scale, with redder colors indicating higher values. The levels of the color scale differ from panel to panel, as they have been set automatically to ensure an adequate sampling of the dynamical range of the surface density shown in each map. Spatial scales in both axes are provided in simulation units. The radial extent of each independent IC (i.e., characterized by an unique exponential profile, see \S\ref{sec:sbr}) are marked with dashed lines in each map. }\label{fig:mosaicmorph1}
\end{center}
\end{figure*}

\begin{figure*}[t]
\begin{center}
\includegraphics*[width=\textwidth]{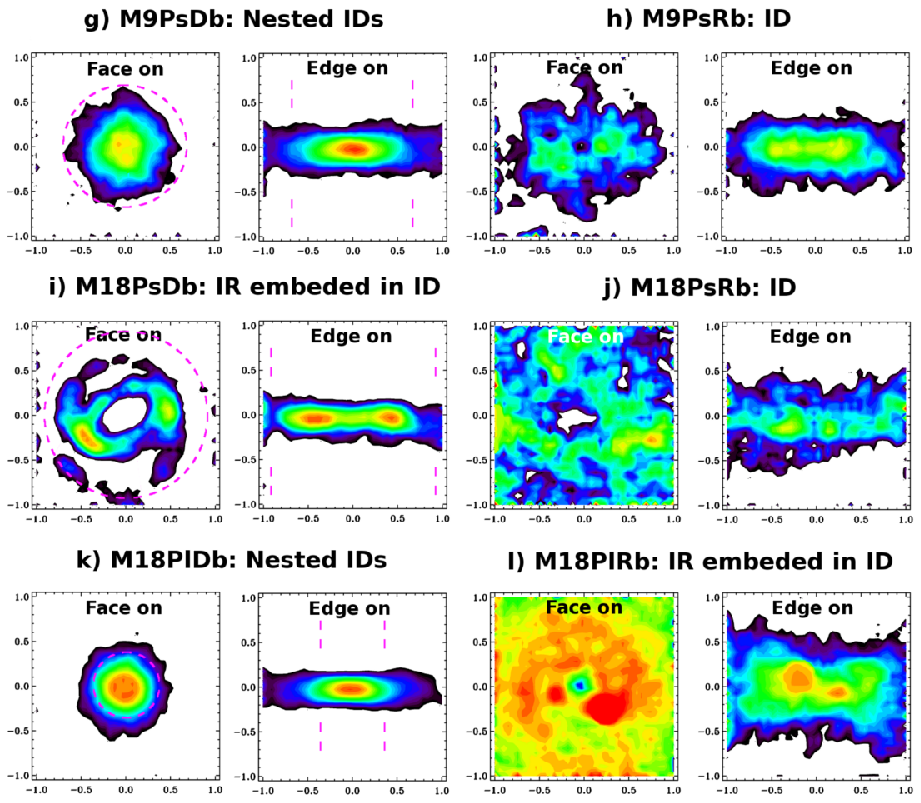}
\caption{Morphology of the ICs resulting in models with mass ratio 1:9 and 1:18 out of disrupted satellite stellar material (models g-l in Table\,\ref{tab:models2}). See caption of Fig.\,\ref{fig:mosaicmorph1}.}\label{fig:mosaicmorph2}
\end{center}
\end{figure*}

\begin{figure}[t]
\begin{center}
\includegraphics[width=0.5\textwidth]{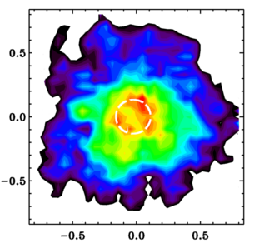}
\caption{Pseudo-ring structure in the final remnant of model M9PsRb, made of stars that initially belonged to the primary disc (experiment h in Table\,\ref{tab:models2}). It coincides with the location of the two twin symmetric clumps observed in the stellar satellite material in the final remnant. Both substructures are associated to an oval distortion induced by the satellite accretion (see Fig.\,\ref{fig:mosaicmorph2}). The pseudo-ring has been marked with a white dashed circle. A rainbow colour palette is used to represent different surface density levels in logarithmic scale, with redder colors indicating higher values. All physical quantities are given in simulation units.}\label{fig:irprimary}
\end{center}
\end{figure}

The final distribution of primary disc material does not show distinct ICs, except in experiment M9PsRb (see Fig.\,\ref{fig:irprimary}), where the primary disc material develops a pseudo-ring at the location of the twin clumps observed in the IC made out of satellite material (see panel h in Fig.\,\ref{fig:mosaicmorph2}). Therefore, we are led to conclude that the minor mergers do not drive the formation of IDs or IRs made out of primary material in our experiments, but force the introduction of primary disc material to the centre instead, making the final galaxy bulge to grow larger (as reported in EM06). 

The resonances that the minor merger induce in the primary disc couple with the satellite disruption, triggering the formation of ICs, such as IDs and IRs (see panels d in Fig.\,\ref{fig:mosaicmorph1}, and i in Fig.\,\ref{fig:mosaicmorph2}). Nevertheless, no clear nuclear bars are developed in the IC formed out of disrupted satellite material either. So, our models prove that minor mergers can give place to the formation of IRs without requiring the development of noticeable bars.

\begin{table*}
\caption{Characteristics of the ICs formed in the remnants}
\label{tab:ics}
\centering
\begin{tabular}{lllll}
\hline\hline
\multicolumn{1}{c}{Model} & \multicolumn{1}{c}{Global structure} &  \multicolumn{1}{c}{Substructure} & \multicolumn{1}{c}{Rotation} & \multicolumn{1}{c}{Detectable features}  \\
\multicolumn{1}{c}{(1)}& \multicolumn{1}{c}{(2)}  & \multicolumn{1}{c}{(3)}    & \multicolumn{1}{c}{(4)}  & \multicolumn{1}{c}{(5)}   \\\hline\vspace{-0.3cm}\\
(a) M6 Ps Db & Nested IDs (thin and thick) & Clumps     & Co-rot. & --- \\
(b) M6 Ps Rb & Nested IDs (both thick)     & Oval distortion      & Counter-rot. & $h_4>0$ at the ID location. \\
(c) M6 Pl Db & Nested IDs (both thin)      & Clumps  & Co-rot. & $\epsilon$ and PA trends of weak oval distortion. \\
(d) M6 Pl Rb & IR$+$ID (thick)             & Clumps  & Counter-rot. & $\epsilon$ and PA trends of weak oval distortion.\\
              &                &    &      & Ring sections in $h_3$ and $h_4$ maps.\\
              &                &    &      & Dumbbell structure in $\sigma$ map.\\
(e) M6 Ps Ds & Bulge$+$ID (thick)          & ---           & Co-rot.  & --- \\
(f) M6 Ps Rs & Bulge$+$ID (thick)          & ---           & Counter-rot. & Central dip in $\sigma$ map.\\
              &                &    &      & Dumbbell structure in $\sigma$ map. 
\vspace{0.05cm}\\\hline\vspace{-0.3cm}\\
(g) M9 Ps Db &  Nested IDs (thin and thick)& Clumps & Co-rot. & $h_4>0$ at the ID location. \\
(h) M9 Ps Rb &  ID  (thin)                 & Clumps  & Counter-rot. & --- 
\vspace{0.05cm}\\\hline\vspace{-0.3cm}\\
(i) M18 Ps Db &  IR$+$ID (thin)            & Clumps, spiral arms & Co-rot. &  $h_4>0$ at the ID location.\\
(j) M18 Ps Rb &  ID (thin)                 & Irregular clumpy structure & Counter-rot. & Central dip in $\sigma$ map.\\
(k) M18 Pl Db &  Nested IDs (thin)         & Ring relics & Co-rot. & --- \\
(l) M18 Pl Rb &  IR$+$ID (thin)            & Pseudo-ring & Counter-rot.  & --- \\\hline\\
\end{tabular}
\begin{minipage}[t]{\textwidth}\vspace{0.1cm}
\emph{Columns}: (1) Model code. (2) Global structure of the formed IC in each remnant, as derived from the radial and vertical surface density profiles shown in Figs.\,\ref{fig:sbrs2}-\ref{fig:zscale} (\S\S\ref{sec:sbr}-\ref{sec:hz}). (3) Substructure found in the face-on surface density maps of Figs.\,\ref{fig:mosaicmorph1}-\ref{fig:mosaicmorph2} (\S\ref{sec:zoo}). (4) Rotation sense of the IC with respect to the global rotation pattern of the remnant (co- or counter-rotating), as obtained in \S\ref{sec:rotation}. (5) Special detectable features present in the global ellipticity and position angle (PA) profiles and in the global kinematical maps of the final remnant, pointing to the existence of these ICs. All the ICs exhibit abrupt changes of the trend in $\epsilon$ and PA profiles at the limits of the ICs, as well as rotation in the center, so none of these two features are listed in the Table (see Figs.\,\ref{fig:mosaico1}-\ref{fig:mosaico4} in \S\ref{sec:detection}).
\end{minipage}
\end{table*}

\subsubsection{Alignment of ICs in the remnants}
\label{sec:misalignment}

\begin{figure}[t]
\begin{center}
\includegraphics[width=0.5\textwidth]{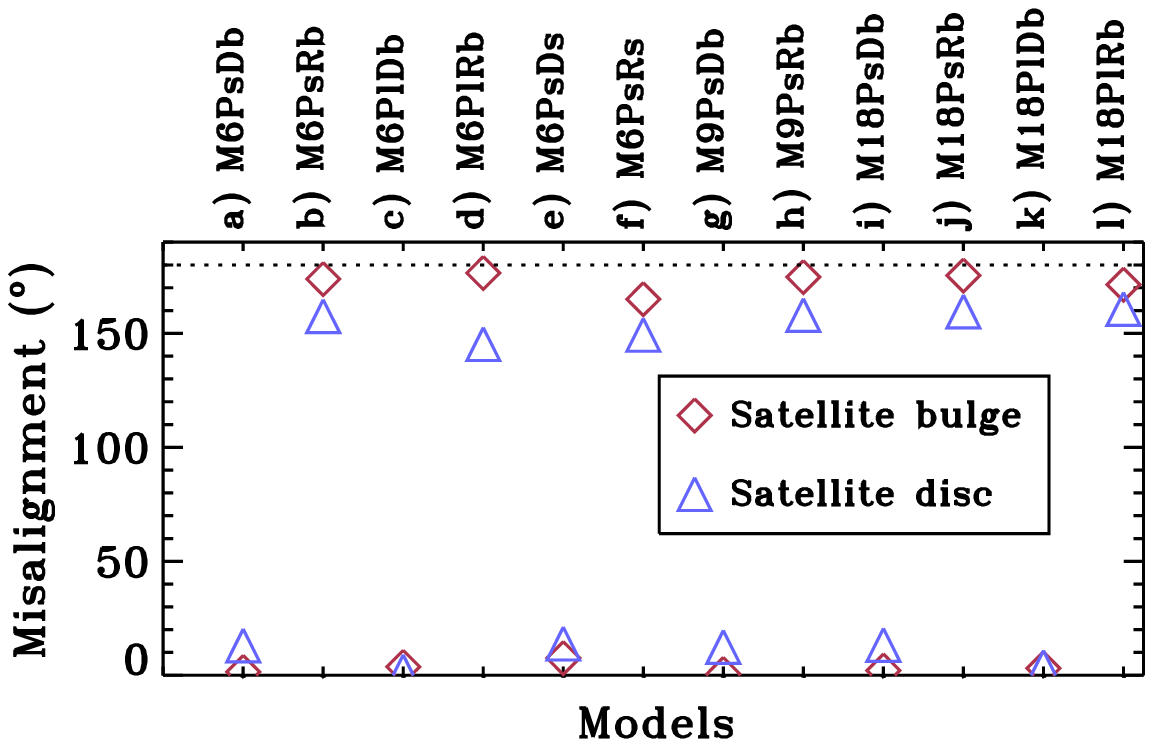}
\caption{Misalignment between the material initially belonging to the satellite bulge and disc with respect to the luminous galactic plane of the final remnants (diamonds and triangles, respectively). The letters identifying each model are those used in Table\,\ref{tab:models2}.}\label{fig:misalignment}
\end{center}
\end{figure}

We have analysed the alignment of the different inner structures formed in the merger with respect to the global remnant structure. This analysis is restricted by the fact that our models explore only two orbit inclinations (30$\deg$ and 150$\deg$), hence little can be said on the effect of different inclinations in the alignment of the resulting ICs. In Fig.\,\ref{fig:misalignment}, we plot the angles between the angular momenta of the stars originally belonging to the satellite bulge and disc in the final remnant and its total angular momentum (diamonds and triangles, respectively). This figure shows that there is a nearly co-planarity between the IC formed out of satellite bulge material and the final galactic plane of the remnant in all the models (notice that their angles are $\sim 0\deg$ in the direct orbits and $\sim 180\deg$ in the retrograde ones). The alignment of the structure formed out of satellite disc material is slightly lower. The near perfect alignment indicates that the plane of the orbit of the satellite has precessed to that of the primary disc by the time the satellite nucleus disrupts, but that precession is incomplete when the satellite loses its disc.

The rotation sense of the IC resulting from the disrupted satellite material can be deduced from this figure: it rotates in the same direction as the main galaxy disc in direct orbits and counter-rotates in the retrograde cases. In general, the alignment between the formed ICs and the main body of the galaxy is worse in retrograde experiments \citep[as already known, see references in][]{2009arXiv0909.1812B}. Nevertheless, we can conclude that the ICs resulting in these experiments are highly aligned with the final galactic plane in all the cases. 

\subsubsection{Rotational support of the ICs}
\label{sec:rotation}

The ICs formed in our experiments are strongly rotationally-supported, with $V_\mathrm{max}/\sigma \sim 2.5$ typically. In Fig.\,\ref{fig:rotcurve}, we show the final velocity and velocity dispersion 2D-maps of models M9PsDb and M9PsRb (models g and h in Table\,\ref{tab:models2}), considering all the stars in the remnants, the stars from the primary galaxy, those from the satellite disc, and those from the satellite bulge. The isophotes of the material considered in each panel are also shown, as well as its photometric and kinematic axes. The rotation velocity maps of both models show that the global velocity field of the remnant is governed by the material of the primary galaxy. In fact, the existence of a co- (or counter-) rotating IC made out of satellite material in the remnant core affects negligibly to the iso-velocity contours of the global maps (notice that the maps considering all the stars and only those from the primary galaxy in the remnants are extremely similar in both models). Moreover, the existence of these ICs does not affect appreciably to the velocity dispersion maps either (compare the $\sigma$ maps of each model including and excluding the satellite material). 

Nevertheless, the ICs affect noticeably to the orientation of the photometric axes with respect to the kinematic ones. As shown in the figure, those obtained just considering the primary galaxy material are noticeably rotated with respect to the ones obtained considering all the stellar content (compare the white axes drawn in the two first velocity maps of the direct model). This is because direct minor mergers drive the formation of large warps in the primary disc \citep[][]{2010arXiv1006.1659R,2010arXiv1006.4855S}. The isophotes in the remnant outskirts of the primary disc material trace these warps, rotating the corresponding photometric axes with respect to the orientation (see Fig.\,\ref{fig:warp}). However, the addition of a highly-aligned IC in the centre increases the weight of the central regions in the photometry, making the kinematic and photometric axes to be more similar (compare the kinematic and photometric axes in the first velocity map of the direct model in Fig.\,\ref{fig:rotcurve}). 

On the other hand, a tight alignment between the photometric and kinematic axes is observed in retrograde mergers, independently on whether we consider the satellite stars in the maps or not. The reason is that retrograde orbits give place to much weaker warps than direct ones due to their lower spin-orbit coupling (compare the velocity maps of the primary disc material in both models of the figure). 

The satellite material in the retrograde models counter-rotates with respect to the material originally from the primary disc (compare the velocity maps of the retrograde model in Fig.\,\ref{fig:rotcurve}). However, no counter-rotation is imprinted to the material coming from the primary galaxy in any model (but see \S\ref{sec:bender}). The existence of counter-rotating ICs contribute to rise the velocity dispersion in the remnant centre, but it does not imprint any significant counter-rotation in the global velocity field of the remnants. Similar conclusions can be derived from analogous maps to the ones shown in Fig.\,\ref{fig:rotcurve} in the other models.

The findings of \S\S\ref{sec:sbr}-\ref{sec:rotation} shows that minor mergers can give place to highly-aligned rotationally-supported ICs out of accreted satellite material, without requiring a noticeable re-distribution of primary galaxy material through a strong bar or relevant dissipative processes (see \S\ref{sec:discussion}). The rotation sense of each IC with respect to the global rotation field of the final remnant for each model is indicated in Col.\,4 of Table\,\ref{tab:ics}.

\subsubsection{Bender diagrams}
\label{sec:bender}

Bender diagrams of our remnants have been obtained using the procedure introduced by \citet{2006MNRAS.372L..78G}. We obtain line-of-sight velocity distributions (LOSVD) for the merger remnants by choosing a point of view at random for projecting the particle distribution. We then derive a surface density map and define iso-density contours to fit ellipses deriving values for the ellipticity, PA, and the $a_4$ parameter (the fourth-order Fourier coefficient measuring the deviation of the iso-density contours from pure ellipses). We place a slit along the major axis of our ellipses up to $R = 2$. Given the primary disc scale lengths, our mapping reaches well into the region of the disc. We have binned the slit in 10 spatial bins and the velocity interval in 50 bins. We find the radial projected velocity and the number of particles in each bin in velocity for each bin in the slit. In this way we obtain a line-of-sight velocity distribution. Finally, we fit the LOSVD by a Gaussian and the residuals by a Gauss-Hermite polynomial as given by \citet{1993ApJ...407..525V} and \citet{1994MNRAS.269..785B}. We repeat this process for each remnant for 90 randomly chosen points of view.

From the fitting procedure we obtain for each LOSVD a value for the velocity centroid of the distribution at each spatial bin along the slit ($V$), the velocity dispersion ($\sigma$), and the amplitude of the third Hermite polynomial ($h_3$), which is a measurement of the skewness of the distribution. 

We show the resulting Bender diagrams for our simulations with 1:6 mass-ratio in Fig.~\ref{fig:bender} (the results from the other simulations behave similarly). All models present an anti-correlation between the $V/\sigma$ and $h_3$ parameter. This is something to be expected due to the prominence of short-axis tube orbits in the progenitor discs that are kept relatively undisturbed due to the merging event. The large bulge of simulations M6PsDb, M6PsRb, M6PlDb, and M6PlRb (labelled from a to d in Table\,\ref{tab:models2}) also helps stabilizing these orbits against the effects of the collisionless minor merger. These simulations differ in the amplitude of the skewness parameter or the $V/\sigma$ as a consequence of the different orbital parameters of the encounters. 

An interesting difference is to be observed in the two plots corresponding to simulations with small bulges in the primary disc: M6PsDs and M6PsRs (models e and f in Table\,\ref{tab:models2}). Here most of the LOSVD present the anti-correlation usual for discs. However for small values of $V/\sigma$ we find a mild (in the case of M6PsDs) or stronger (in M6PsRs) correlation of the two parameters. This is due both to the effect of the small bulge that is not able to stabilize short-axis tube orbits and to the larger contribution of the satellite bulge in the inner parts of these simulations. It is interesting to note that the effect of the counter-rotating orbit of M6PsRs is signalled by the relatively large correlation signature in the central parts of the diagram.

\subsection{Detection of ICs in the global stellar maps of remnants}
\label{sec:detection}

The prominence of the primary stars in the final remnant make the ICs to be completely masked by their mass distribution, in such a way that no direct detection of any IC can be performed in either the global surface density maps and profiles of the final remnants. It is encouraging though that indirect detection similar to that attempted and used in observations proves to be successful in many cases, as we will show in this section.

\subsubsection{Ellipticity and position angle profiles of the remnants}
\label{sec:ellipticity}

Inner structures in real galaxies imprint noticeable features in the ellipticity ($\epsilon$) and PA profiles of the central isophotes of the host galaxies \citep[][]{2003ApJ...597..929E,2004MNRAS.350...35F,2009A&A...499L..25C}. In the right panels of  Figs.\,\ref{fig:mosaico1}-\ref{fig:mosaico4}, we have plotted the $\epsilon$ and PA profiles of the stellar material in all the remnants, assuming an inclined viewing angle ($\theta=60\deg$, $\phi=20\deg$). The radial extent of the ICs as defined in \S\ref{sec:sbr} have been marked in all the  plots (vertical dashed lines).

The profiles of the remnants with big primary bulges are dominated by the light distribution of the pre-existing primary bulge in the center, leading to low ellipticities in the remnant even at the presence of ICs ($\epsilon \leq 0.2$, see the cases shown in Fig.\,\ref{fig:mosaico1}, for example). Nevertheless, the existence of ICs (and even their radial extensions) can be deduced just attending to the abrupt changes in the trends of the $\epsilon$ and PA profiles that appear at the transition region between adjacent ICs (for example, between two nested IDs or between an ID and the outer remnant disc, see Figs.\,\ref{fig:mosaico1}-\ref{fig:mosaico4}). These trend changes are also observed in real galaxies, associated to isophotes twisting produced by bars and triaxial structures in the bulges \citep[see][]{1997A&AS..125..479J,2003ApJ...597..929E}. 

The existence of weak bar-like or oval distortions in the remnant center of some models can be deduced from their global isophotal profiles. Models M6Pl[D/R]b (panels c in Fig.\,\ref{fig:mosaico1} and d in Fig.\,\ref{fig:mosaico2}) exhibit a slight maximum in $\epsilon$ and a constant PA at the radii where the IC presents twin clumps (see the corresponding panels in Fig.\,\ref{fig:mosaicmorph1}). Nevertheless, as commented above, none of our remnants develop noticeable nuclear bars, a fact that is corroborated by the global ellipticity and PA profiles shown in Figs.\,\ref{fig:mosaico1}-\ref{fig:mosaico4}.

The models with a small primary bulge exhibit central ellipticities equal to or greater than that of the outer disc ($\epsilon \sim 0.3$, see Fig.\,\ref{fig:mosaico2}), suggesting that the inner light is dominated by a disc or flattened triaxial structure \citep{2003ApJ...597..929E}. The profiles of these models exhibit ellipticity and PA profiles typical of bulge$+$disc galaxies. The ellipticity shows a quick rise in the core (bulge) region, a slight decrease in the bulge-disc transition region, and a constant value in the disc outer layers, while the PA is nearly constant outside the bulge region (see Fig.\,\ref{fig:mosaico2}). Some particular features in the $\epsilon$ and PA profiles corroborating the existence of some of the ICs identified in \S\S\ref{sec:sbr}-\ref{sec:zoo} are listed in Col.\,5 of Table\,\ref{tab:ics}.

\begin{sidewaysfigure*}[ht]
\vspace{18cm}
\begin{center}
\includegraphics*[width=0.9\textwidth,angle=0]{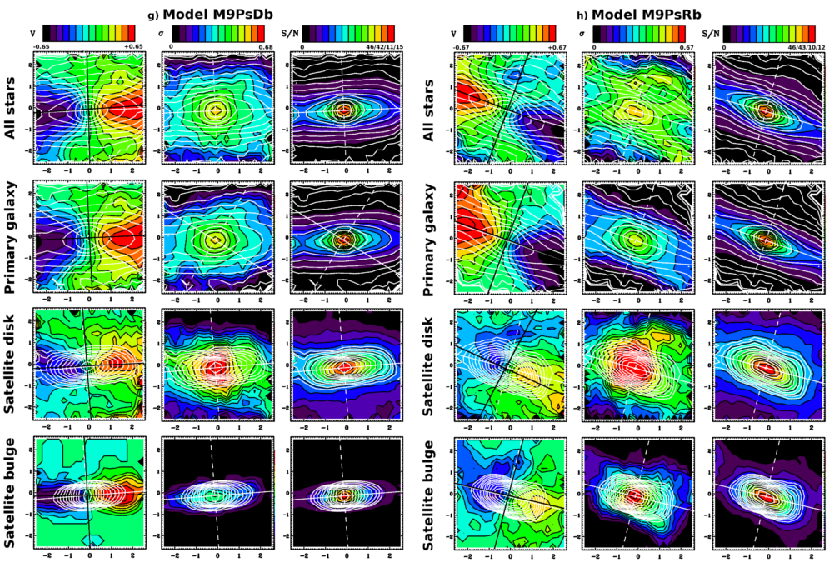}
\caption{Line-of-sight rotation velocity, velocity dispersion, and signal-to-noise ratio ($S/N$) maps of the stellar material coming from different original components in the final remnants of models M9PsDb and M9PsRb, using an edge-on view of the remnants (experiments g and h in Table\,\ref{tab:models2}). The same levels have been used for the color palette of the velocity maps of the different components in each model and in the velocity dispersion maps to remark the high rotational support of the ICs resulting from satellite material (consult the bars on the top of the corresponding columns). Levels are distributed linearly. The maximum level in the bars corresponding to the $S/N$ maps show four different values, corresponding to each one of the frames below, respectively. All physical magnitudes are provided in simulation units. \emph{First row}: Considering all the stars in the remnant. \emph{Second row}: For the stars initially belonging to the primary galaxy. \emph{Third row}: For the stars originally in the satellite disc. \emph{Fourth row}: For the stars originally in the satellite bulge. \emph{Contours}: Surface iso-density contours in the final remnant of the material considered in each map, just as reference. \emph{White straight lines}: Photometric axes of the material considered in each panel. \emph{Black straight lines}: Kinematical axes of the material considered in each panel.}\label{fig:rotcurve}
\end{center}
\end{sidewaysfigure*}

\subsubsection{Kinemetric moments of the velocity fields in the remnants}
\label{sec:vmap}

In many real galaxies, the existence of kinematically-decoupled components that are not detectable through direct imaging is inferred from special features in their kinematic maps. In most cases, the existence of these ICs is later confirmed by unsharp masking or by direct detection of its gas component in [OIII] or H$\beta$ emission-line maps \citep[see, e.g.,][F06]{2003A&A...405..455F}. 

Figures\,\ref{fig:mosaico1}-\ref{fig:mosaico4} show the 2D-maps of the kinemetric moments of the line-of-sight velocity distribution of all the luminous material in each remnant, using edge-on views. The line-of-sight velocity $V_\mathrm{LOS}$, the velocity dispersion $\sigma$, and the third- and fourth-order coefficients of the Gauss-Hermite expansion $h_3$ and $h_4$ are shown for each model. They have been obtained adapting the \emph{profit} routine (originally designed for profile fitting of spectral emission-lines by Gauss–Hermite series) to N-body data \citep[][]{2010Ap&SS.327..239R}. The rotation velocity maps in these figures differ from the analogous ones presented in Fig.\,\ref{fig:rotcurve} in the spatial resolution. While those presented in Fig.\,\ref{fig:rotcurve} use an uniform spatial binning, the ones shown in Figs.\,\ref{fig:mosaico1}-\ref{fig:mosaico4} have higher spatial resolution in the center than in the outskirts to improve the $S/N$ of the estimates of the kinemetric moments in low-density regions (analogously to what is done in observations, see F06). The central spatial resolutions in these figures are similar to those achieved by current observations in the nearby Universe, adopting the scalings proposed in \S\ref{sec:models} ($\sim 0.5$\,kpc). Regions with low $S/N$ have been masked in the maps.

Figures\,\ref{fig:mosaico1}-\ref{fig:mosaico4} show that the existence of many ICs could be deduced from these maps of kinemetric moments. Many particular features present in these maps that corroborate the existence of some of the ICs identified in \S\S\ref{sec:sbr}-\ref{sec:zoo} are indicated in Col.\,5 of Table\,\ref{tab:ics}. However, we must remark that these features are smooth and show large dispersion (specially, at outer radii), and hence, they could be detectable only in case of low noise levels. Some of these kinematical features include:

\indent {\bf i. Misalignment of kinematic axes}.

\begin{figure}[t]
\begin{center}
\includegraphics[width=0.5\textwidth]{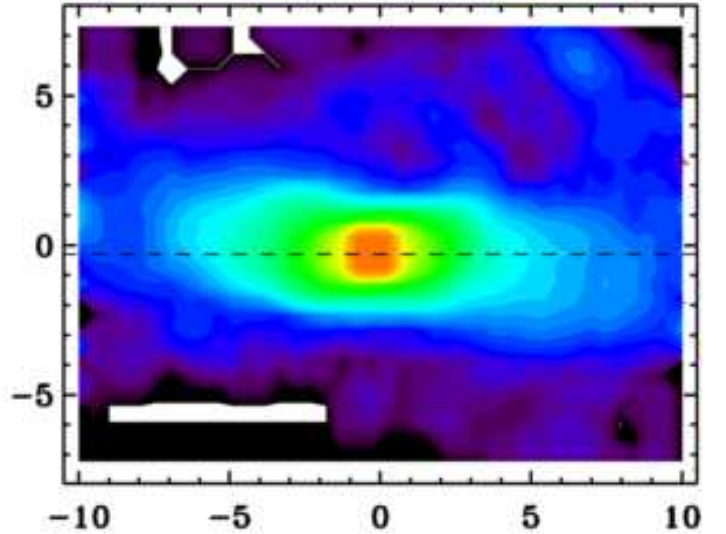}
\caption{Final warped disc of model M9PsDb (experiment g in Table\,\ref{tab:models2}). An edge-on view of the material originally belonging to the primary disc in the final remnant of this model is shown. A rainbow colour palette is used to represent different surface density levels in logarithmic scale, with redder colors indicating higher values. All physical quantities are given in simulation units. }\label{fig:warp}
\end{center}
\end{figure}

A clear misalignment of the inner and outer kinematic axes of the galaxy is detected in some cases (see model M6PsRb, panel b in Fig.\,\ref{fig:mosaico1}), which can be produced by noticeable disc warping (as the case shown in Fig.\,\ref{fig:warp}). Rotation reaches the central regions in all the remnants. The final rotation fields, although following a typical spider-like diagram, seem quite distorted in general, as it is characteristic of dissipationless models \citep{2007MNRAS.376..997J}. In general, experiments with lower mass ratios or larger pericenters imply a smoother destruction of the initial galaxy rotation pattern. \\[-0.3cm]

\indent {\bf ii. Stretching of iso-velocity contours in the centre}.

The existence of co-rotating IDs and IRs imprints higher rotation in the centre, stretching the iso-velocity contours towards the major axis at their location. This makes the angle of these contours to be more open at the remnant outskirts than in the centre (see, \eg, models M6PsDs and M6PlDb, panels b and c in Fig.\,\ref{fig:mosaico1}). \\[-0.3cm]

\indent {\bf iii. Twisting of central iso-velocity contours in retrograde models.}

As shown in \S\ref{sec:rotation}, retrograde models give place to counter-rotating ICs. In some of them, the contribution of this IC to the mass in the remnant core is high enough to imprint a strong twisting of the iso-velocity contours by $\sim 90\deg$ at the core region (see model M6PsRb in panel b of Fig.\,\ref{fig:mosaico1}, and models M18P[s/l]Rb in panels j and l of Fig.\,\ref{fig:mosaico4}).\\[-0.3cm]

\indent {\bf iv. S-shaped twists of iso-velocity contours.}

Almost all the cases exhibit S-shaped or integral-sign-shaped twisting of the central iso-velocity contours at a certain height in the galactic plane. These features are produced by the sharp decrease of the amount of stars that contributes to the velocity field at a certain spatial position. Therefore, S-shaped kinematic twists in our models appear at the spatial locations where the mass contribution of the ICs becomes negligible (see, \eg, models M6PlRb and M18PsDb, in panels d and i of Figs.\,\ref{fig:mosaico2} and \ref{fig:mosaico3}).\\[-0.3cm]

\indent {\bf v. $h_4$ peaks.}

The majority of the models with big primary bulges show positive $h_4$ values in general, peaking at the location of the ICs formed at the end of the simulation. In particular, the ring geometry of the IC resulting in model M6PlRb can be deduced from the two well-defined symmetric peaks present in the $h_4$ maps, which trace its perpendicular sections (panel d in Fig.\,\ref{fig:mosaico2}).

\indent {\bf vi. Correlations between the kinemetric moments.}

We have found that $v_\mathrm{LOS}$ and $h_3$ correlate at the location of the formed IC in our remnants if it exhibits oval or bar-like distortions (see models M6Ps[D/R]b and M6PlDb in panels a-c of Fig.\,\ref{fig:mosaico1}, and models M9PsRb and M18PsDb in panels h-i of Fig.\,\ref{fig:mosaico3}). However, when the ICs show no trace of having had them, $v_\mathrm{LOS}$-$h_3$ anti-correlate at the location of the IC, accordingly to the discy structure of the majority of the obtained ICs (see models M18PsRb and M18Pl[D/R]b in Fig.\,\ref{fig:mosaico4}). Some models also exhibit a mixed behaviour:  $v_\mathrm{LOS}$-$h_3$ correlate at the remnant core (at $R<0.3$) and anti-correlate at the position of the $h_4$ peaks, as occurs in models M6PsDs (panel e in Fig.\,\ref{fig:mosaico2}) and M9PsDb (panel g in Fig.\,\ref{fig:mosaico3}). This is probably pointing to an origin related to oval distortions and bar-like distortions in these components.

\indent {\bf vii. Dumbbell $\sigma$ structures.}

Dumbbell structures in $\sigma$ maps point to the existence of counter-rotating IDs, which rise $\sigma$ at their location. Although all the retrograde models give place to counter-rotating IDs or IRs, dumbbell structures are only observed in models M6PlRb and M6PsRs (panels d and f in Fig.\,\ref{fig:mosaico2}).  \\[-0.3cm] 

\indent {\bf viii. Central $\sigma$ dips.}

In general, the velocity dispersion is very high in the remnant centres because the bulge dynamics dominate at those radii. However, it tends to decrease at the bulge-to-disc transition region due to the contribution of the co-rotating IC. In our models, $\sigma$ dips are detected only in the models of small primary bulges (see panels e and f of Fig.\,\ref{fig:mosaico2}). \\

Summarizing, the velocity fields of our remnants exhibit features that reveal the existence of the central ICs formed through the satellite accretion, such as the stretching of the iso-velocity contours in the centre, S-shaped kinematic twists, and dumbbell $\sigma$ profiles. More inclined views of the remnant smooth out these features, making the ICs undetectable in the velocity 2D-maps. 

\section{Qualitative comparison to observations}
\label{sec:observations}

In this section, we make a qualitative comparison of the properties of the ICs obtained in our minor merger simulations with real observational cases, to stress the similarities and differences among them. 

\subsection{Structural comparison}
\label{sec:observationsmorph}

\begin{figure*}[t]
\begin{center}
\includegraphics*[width=\textwidth]{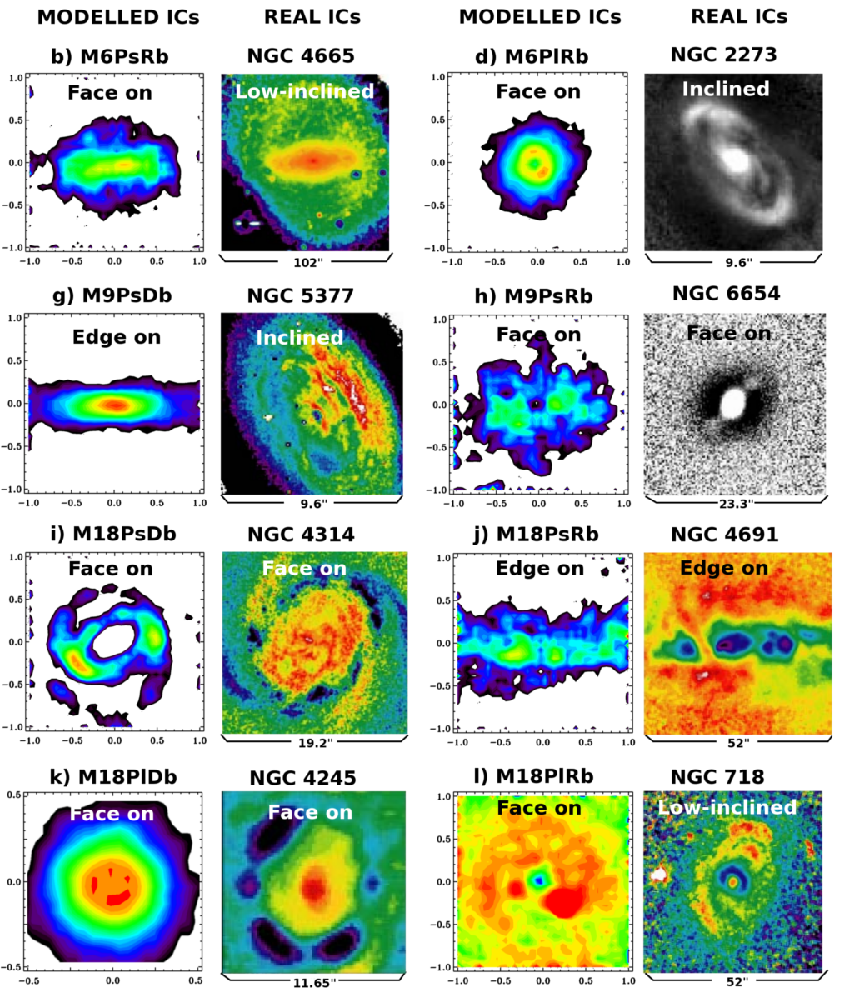}
\caption{Comparison of some ICs obtained in our minor merger experiments (first and third columns in the figure) to real observational examples with similar morphologies (second and fourth columns, respectively). The surface density maps of the ICs generated in our models are taken from Figs.\,\ref{fig:mosaicmorph1}-\ref{fig:mosaicmorph2} and are in simulation units (consult captions there). The observational examples are taken from the sample of spiral galaxies with ICs developed by \citet{2003ApJS..146..299E}. Coloured maps of real galaxies correspond to $B-R$ or $V-I$ color maps of these galaxies, while their grey-scale maps represent unsharp masks in $V$, $R$ or $H$ bands (see \citet{2003ApJS..146..299E} for a detailed description of each frame). The global inclination of galaxy in the observational cases and the spatial scale of each postage stamp are indicated in each frame. In images of real galaxies, North is up and East is left, except in NGC\,4665, which has been rotated 90\deg\ clockwise to emphasize the similarity with the IC resulting in experiment M6PsRb (model b). For a comparison of the scaled sizes of the ICs resulting in our models with those exhibited by real ICs, consult \S\ref{sec:observations}. Material from \citet{2003ApJS..146..299E} in this figure is reproduced by permission of the American Astronomical Society (AAS) and of the original authors.}\label{fig:comparacion}
\end{center}
\end{figure*}

Figure\,\ref{fig:comparacion} compares the morphology of some of the ICs formed in our experiments to observational examples that exhibit similar inner structures from the sample of \citet{2003ApJS..146..299E}. As shown by the figure, the ICs resulting from our minor merger experiments are analogous structural and morphologically to the ICs hosted by many real spiral galaxies.  IDs, nested IDs, rings and pseudo-rings, spiral patterns, and even undisrupted clumps with irregular spatial distributions as those observed in our experiments are found in real spiral galaxies \citep[see also][]{1996FCPh...17...95B,2002AJ....124...65E,2003ApJ...597..929E,2004A&A...415..941E,2010MNRAS.402.1397P,2011arXiv1103.1692S}. 

Star-forming pseudo-rings are substructures usually associated to gas and starbursts in real galaxies \citep[F06;][]{2010MNRAS.402.2140S,2010AIPC.1240..255S}. However, although our models do not consider dissipative components, some remnants exhibit stellar ring relics or pseudo-rings in the centre (see models M18Pl[D/R]b in panels k and l of Fig.\,\ref{fig:mosaicmorph2}). 

Our remnants have not developed significant nuclear bars (although some of them present signs of weak oval or bar-like distortions in the center, see \S\ref{sec:ellipticity}). Then, the IRs and pseudo-rings that appear in some of our models are not directly associated to well-formed bars. Instead, they have formed by the coupling of the satellite disruption to the resonances induced in the primary disc by the encounter. Cases of IRs and sets of concentric rings not associated to bars are found in real galaxies too \citep{1996FCPh...17...95B,2004AJ....127...58M,2006ApJ...642..765K}. So, our models point to the resonance-related origin of IRs, independently on whether these resonances are linked to strong bars or to weak non-axisymmetric perturbations as the ovals found in some of our models (see \S\ref{sec:bars}).

The stellar IDs formed in our experiments have radial lengths ranging from $\sim 0.4$ to $\sim 1$ disc scale-length of the final remnant, which are equivalent to radial lengths spanning from $\sim 360$\,pc to $\sim 2.5$\,kpc, adopting the scaling provided in \S\ref{sec:models}. These values are in excellent agreement with those obtained for the IDs detected in real galaxies \citep[][F06]{2002AJ....124...65E,2006MNRAS.366.1151S,2010MNRAS.402.2140S,2011arXiv1103.1692S}. The IRs and pseudo-rings formed in our simulations are embeded in these IDs, and thus, exhibit typically lower linear sizes. Adopting the scalings of \S\ref{sec:models}, our IRs match pretty well the standard 1 kpc-radius IRs that lie between the inner Lindblad resonances of many disc galaxies \citep[]{2002AJ....124...65E,2010MNRAS.402.2140S}. 

Concerning to their vertical distributions, the flared vertical structure of most the single IDs and nested IDs resulting in our models (see Fig.\,\ref{fig:zscale}) is extremely similar to the one derived recently for the galaxy NCG\,7217, composed by two large scale nested stellar discs \citep{2011arXiv1103.1692S}.

The existence and extension of many real ICs are usually derived from abrupt changes in the trends of the ellipticity and PA profiles of the global galaxy isophotes \citep{2002AJ....124...65E,2003ApJS..146..299E,2004A&A...415..941E}. As we have shown in \S\ref{sec:ellipticity}, all the ICs formed in our models imprint similar features in the global profiles of the final remnants, coherently with observations \citep{2001A&A...367..405P,2003ApJS..146..299E}.

The low misalignment of the ICs obtained in the simulations with respect to the global galaxy plane (see \S\ref{sec:misalignment}) is in excellent agreement with observations: F06 report misalignments of typically $\lesssim 20\deg$ for the ICs harboured by Sa-Sb galaxies, and \citet{2011arXiv1102.3801K} find that 90\% of the ICs in a sample of 260 early-type galaxies can be considered aligned to better than 5 deg. Nevertheless, we can not discard the possibility that this may be facilitated by the moderate inclinations of our simulated orbits. However, the quick decay of the satellite orbit to the primary disc plane observed in all the models suggests that more inclined orbital configurations could result into aligned ICs too, and that extreme orbital inclinations would be required to produce noticeably misaligned ICs. 

Our simulations also demonstrate the feasibility of the scenario proposed by F06 to explain the low misalignment of the ICs observed in Sa-Sb galaxies as compared to those in E-S0's \citep[usually above 60\deg , see][]{2006MNRAS.366.1151S}. According to these authors, if the ICs have resulted from minor mergers, the morphology of the primary galaxy in the encounter should play a crucial role in determining its alignment. In our simulations, the evolution of the satellite orbit to the primary disc plane prior to its disruption is produced by the low spheroidality of the primary galaxy potential, which establishes a privileged plane a priori (the primary disc plane). Spheroidal potentials (as those of E-S0 galaxies) do not have any privileged direction, and thus they do not produce this effect in the orbits of the accreted satellites \citep{2008A&A...477..437D}. Therefore, our simulations suggest that, if the majority of ICs derive from minor mergers, we should expect to detect more aligned ICs in disc galaxies (as Sa-Sb's) than in spheroidal ones (E-S0's), coherently with observations and with F06 arguments. Moreover, F06 propose that the existence of large gas amounts involved in the merger must contribute to this alignment. Notice that, although the previous sentence must be true, our models prove that gas is not strictly necessary to obtain high co-planar ICs in minor mergers.

\subsection{Kinematical comparison}
\label{sec:observationskinem}

ICs disturb the velocity field of their host galaxies, giving place to particular kinematical features in the global maps of the galaxy pointing to their existence. Our remnants exhibit many of these features, such as disturbed iso-velocity contours in the center, noticeable S-shaped kinematic twists at the limiting radii of the ICs, $\sigma$ peaks, dumbbell-like $\sigma$ structures associated to counter-rotating IDs, and stretching of iso-velocity contours at the presence of co-rotating ICs (see \S\ref{sec:rotation}), similarly to what it is observed in real galaxies harbouring ICs \citep[see F06;][]{2006MNRAS.366.1151S,2008MNRAS.390...93K}.

\begin{figure*}[t!]
\begin{center}
\includegraphics*[width=0.3\textwidth]{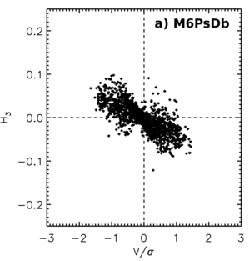}
\includegraphics*[width=0.3\textwidth]{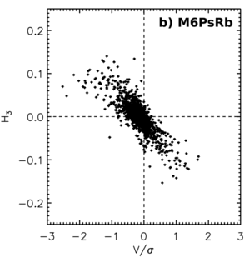}
\includegraphics*[width=0.3\textwidth]{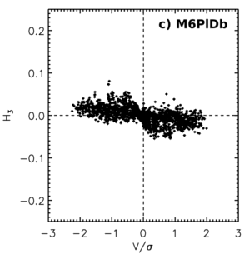}
\includegraphics*[width=0.3\textwidth]{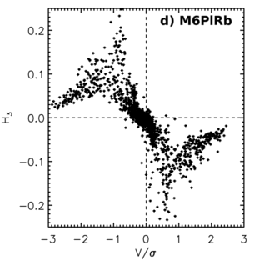}
\includegraphics*[width=0.3\textwidth]{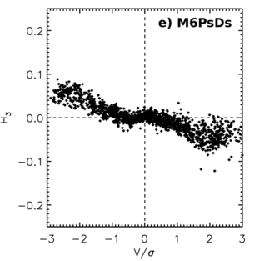}
\includegraphics*[width=0.3\textwidth]{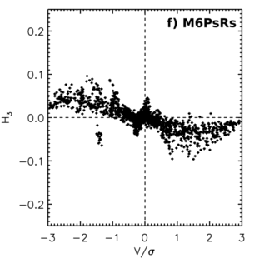}
\caption{Bender diagrams of the remnants resulting from the minor merger experiments run with a mass ratio 1:6. The amplitude of the third Hermite polynomial $h_3$ of the final remnant is plotted against its $V/\sigma$ for 90 randomly chosen points of view.}\label{fig:bender}
\end{center}
\end{figure*}

The counter-rotating ICs formed in our models are so embeded in the primary disc stellar material that, in general, they do not imprint noticeable counter-rotation at the centre of the remnant (\S\ref{sec:vmap}). The masses of the ICs are too low to affect significantly to the central dynamics of the remnants, which are basically dominated by the host rotating disc instead (see \S\ref{sec:rotation}), although some counter-rotation becomes detectable at specific viewing angles (see model M6PsRs in panel f of Fig.\,\ref{fig:bender}). Therefore, the presence of large disc components in galaxies makes the observational detection of counter-rotating ICs difficult. This can explain why the observational samples of early-type spirals exhibit an apparent absence of counter-rotating features as compared to the samples of E-S0's, which lack of relevant discs \citep[see][F06]{2001AJ....121..140K}. In fact, the frequency of counter-rotating ICs in E-S0's and in Sa-Sb's could be similar, the discs of Sa-Sb's being responsible of masking them.

\begin{figure*}[t]
\begin{center}
\includegraphics*[width=0.85\textwidth,angle=0]{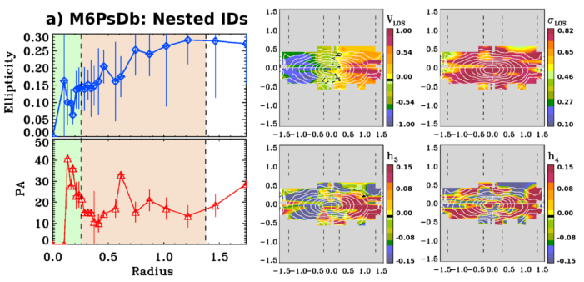}
\includegraphics*[width=0.85\textwidth]{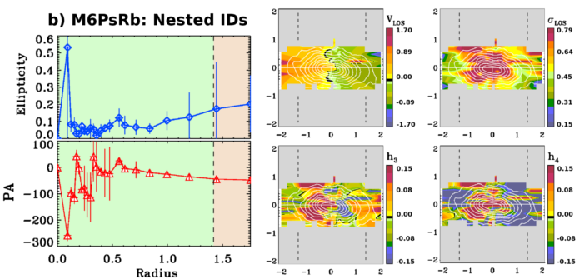}
\includegraphics*[width=0.85\textwidth,angle=0]{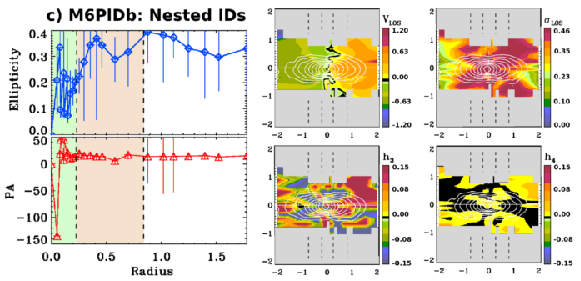}
\caption{Photometrical and kinematical features imprinted by the formed ICs in the global stellar maps of the final remnants of models M6Ps[D/R]b and M6PlDb (models a to c in Table\,\ref{tab:models2}). Left panels: Ellipticity and PA isophotal profiles of all the stars in the remnants, using an inclined view ($\theta=60\deg$, $\phi=20\deg$). The extent of the ICs as defined in \S\ref{sec:sbr} is marked with vertical dashed lines in each case for reference (accordingly to Fig.\,\ref{fig:sbrs2}). Right panels: 2D-maps of the kinemetric moments of the line-of-sight velocity distribution of all the stars in the remnant using an edge-on view ($V_\mathrm{LOS}$, $\sigma$, $h_3$, $h_4$). The levels are distributed linearly (simulation units). The photometric axes defined by the material in the core region of the galaxy are also plotted (\emph{straight lines}). The isophotes of the stellar material in the remnant originally belonging to the satellite are overplotted just as reference (\emph{contours}). }\label{fig:mosaico1}
\end{center}
\end{figure*}

\begin{figure*}[t]
\begin{center}
\includegraphics*[width=0.85\textwidth]{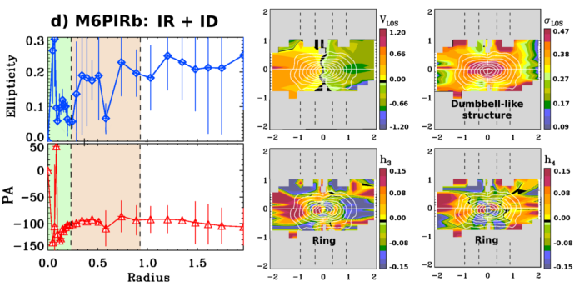}
\includegraphics*[width=0.85\textwidth,angle=0]{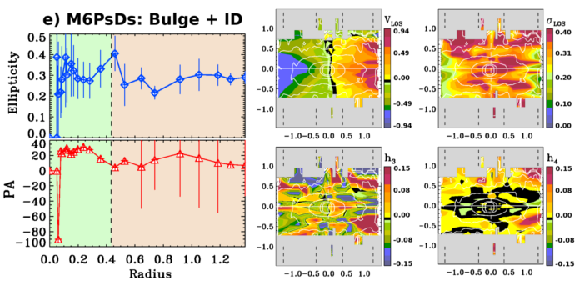}
\includegraphics*[width=0.85\textwidth,angle=0]{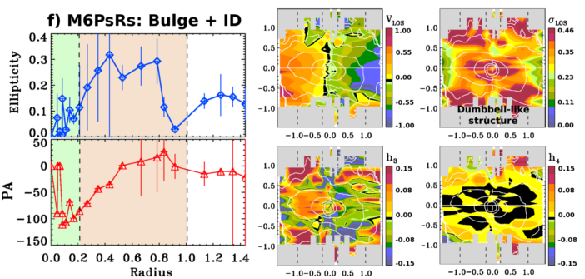}
\caption{Photometrical and kinematical features imprinted by the formed ICs in the global stellar maps of the final remnants of models M6PlRb and M6Ps[D/R]s (models d to f in Table\,\ref{tab:models2}). See caption of Fig.\,\ref{fig:mosaico1}.}\label{fig:mosaico2}
\end{center}
\end{figure*}

\begin{figure*}[t]
\begin{center}
\includegraphics*[width=0.85\textwidth,angle=0]{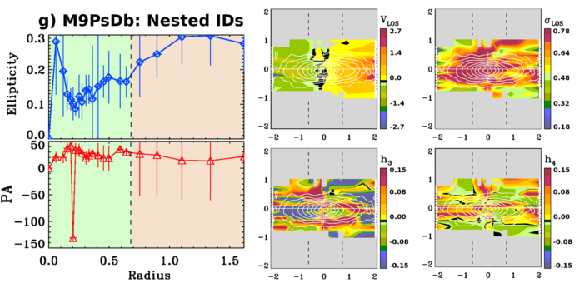}
\includegraphics*[width=0.85\textwidth]{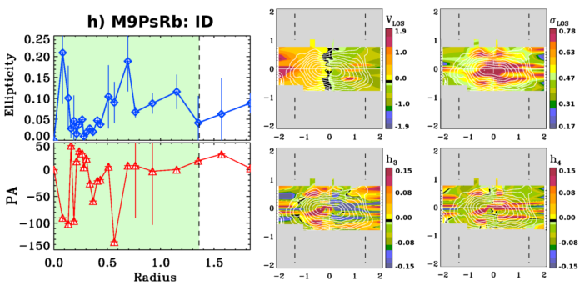}
\includegraphics*[width=0.85\textwidth]{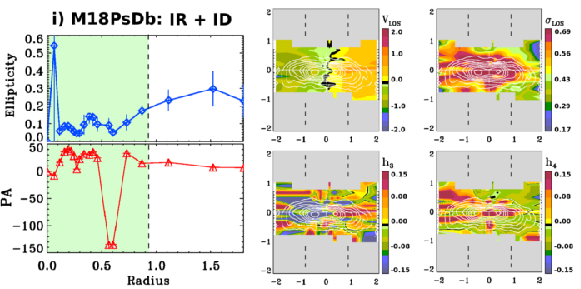}
\caption{Photometrical and kinematical features imprinted by the formed ICs in the global stellar maps of the final remnants of models M9Ps[D/R]b and M18PsDb (models g to i in Table\,\ref{tab:models2}). See caption of Fig.\,\ref{fig:mosaico1}.}\label{fig:mosaico3}
\end{center}
\end{figure*}

\begin{figure*}[t]
\begin{center}
\includegraphics*[width=0.85\textwidth,angle=0]{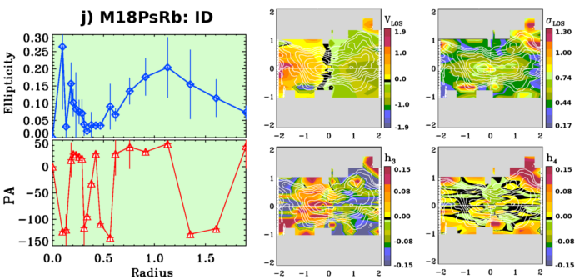}
\includegraphics*[width=0.85\textwidth,angle=0]{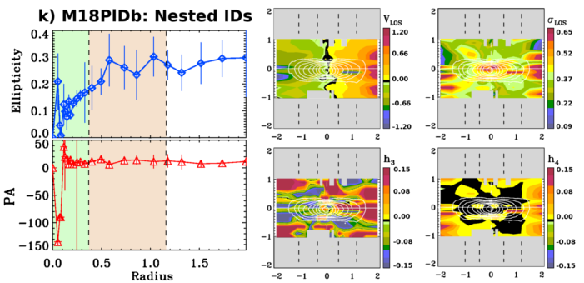}
\includegraphics*[width=0.85\textwidth,angle=0]{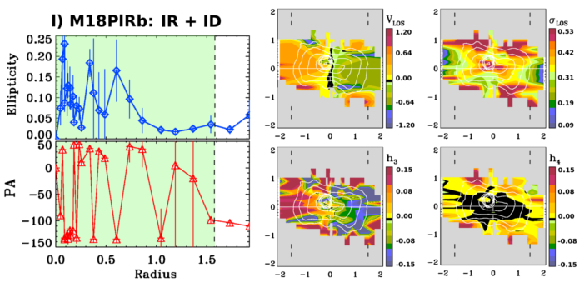}
\caption{Photometrical and kinematical features imprinted by the formed ICs in the global stellar maps of the final remnants of models M18PsRb and M18Pl[D/R]b (models j to l in Table\,\ref{tab:models2}). See caption of Fig.\,\ref{fig:mosaico1}.}\label{fig:mosaico4}
\end{center}
\end{figure*}

Our collisionless models reproduce several trends and correlations between the kinemetric moments of real galaxies with ICs, although they present differences as well. In general, our models show $h_4\geq 0$ and $h_4$ peaks at the location of the ICS, in excellent agreement with observations of ICs in early-type galaxies \citep[F06;][]{2008MNRAS.390...93K}. However, the $V_\mathrm{LOS}$-$h_3$ correlation found in our simulations reproduce the observational trends partially. Observations report $V_\mathrm{LOS}$-$h_3$ anti-correlation at the location of co-rotating ICs in Sa-Sb's, and correlation if the ICs counter-rotate (F06). Nevertheless, our models tend to exhibit $V_\mathrm{LOS}$-$h_3$ anti-correlation, independently on the rotation sense of the IC with respect to the final galaxy (see Figs.\,\ref{fig:mosaico1}-\ref{fig:mosaico4}). Although some retrograde models can present correlation for particular viewing angles (see model M6PsRs, panel f in Fig.\,\ref{fig:bender}), in general we find anti-correlation (see the rest of Bender diagrams in Fig.\,\ref{fig:bender}).

Our models do not include the effects of gas dynamics and star formation, which obviously affect to the kinematical structure of the final remnant (see \S\ref{sec:limitations}). Collisionless remnants are known to trigger box orbits, which move $V_\mathrm{LOS}$ and $h_3$ to correlate \citep[see, e.g.,][]{2001ApJ...555L..91N}. However, the inclusion of gas in the simulation has the opposite effect: it suppresses them and make $V_\mathrm{LOS}$ and $h_3$ to anti-correlate at the remnant centre \citep{2000MNRAS.316..315B,2001ApJ...555L..91N,2007MNRAS.376..997J}. Then, our collisionless simulations behave atypically in this sense, as they tend to exhibit $V_\mathrm{LOS}$-$h_3$ anti-correlation. This trend was already obtained by \citet{2006MNRAS.372L..78G}, who showed that such anti-correlations can be kept in a collisionless merger simulation whenever a central bulge allows the discs to retain some of their original angular momentum during the merger, making short-axis tube orbits to be still present in the final remnant.

In our models, $V_\mathrm{LOS}$-$h_3$ correlation seems to be more related to the existence of  bar-like or oval distortions than to counter-rotation (\S\ref{sec:vmap}). The behaviour observed in our remnants can be reconciled with the observational trends if we consider that retrograde mergers give place to longer-lasting bar-like instabilities (although weaker) than their direct analogues \citep{1996ApJ...471..115B}. Therefore, we should expect to find statistically more bar-like distortions and ovals in the ICs resulting from retrograde mergers than in those coming from a direct encounter. Assuming that $V_\mathrm{LOS}$-$h_3$ correlation is related to the existence of non-axisymmetric distortions in the galaxy disc, the previous result establishes a correspondence between $V_\mathrm{LOS}$-$h_3$ correlations and counter-rotating ICs, coherently with F06 results. This scenario is supported by the different $V_\mathrm{LOS}$-$h_3$ trends exhibited by Sa-Sb and E-S0 galaxies. While ICs hosted by the later ones are always associated to $V_\mathrm{LOS}$-$h_3$ anti-correlations \citep{1994MNRAS.269..785B}, the ICs in early-type discs exhibit correlations only if the IC is counter-rotating (F06). This can be explained considering that E-S0's are more efficient inhibiting bars and ovals than Sa-Sb's (because they have larger bulges and negligible disc components), a fact that should make the ICs in these galaxies to exhibit $V_\mathrm{LOS}$-$h_3$ anti-correlations in general (coherently with observations).

In conclusion, minor mergers could account for the existence of many stellar dynamically-cold ICs in spiral galaxies and even in E-S0's. The present models prove that they can give place to ICs with geometrical, structural, and kinematical properties similar to those observed in real galaxies. 

\section{Model limitations}
\label{sec:limitations}

We have analysed the role of minor mergers in the formation of dynamically-cold thin stellar ICs, without accounting for gas and star formation effects. Obviously, the inclusion of dissipative components in the models would not just provide the formed IC with an additional recent stellar population or gas component, but it could affect noticeably its final structure and kinematics. 

The formation of IDs and IRs are associated to the redistribution of angular momentum in the remnant disc. In this sense, gas components are expected to contribute noticeably to this re-distribution, mainly during disc distortions. Nevertheless, recent studies have demonstrated that dissipative components (although relevant) are not essential or decisive in the formation of kinematically-decoupled ICs through major mergers. Gas causes the remnants to appear more round and axisymmetric, wiping out small kinematical misalignments more easily, but the resulting IC conserves most of its structural properties as compared to collisionless models \citep{2007MNRAS.376..997J}.

Besides, although star formation triggers the formation of young stars in the IC in simulations of major mergers, the resulting IC is still composed by a relevant old stellar component \citep{2008A&A...477..437D}. If the formation of an ID or IR depended basically on merger-induced gas inflows to the galaxy centre and on the subsequent star formation \citep[][]{1996ApJ...471..115B,2005A&A...437...69B}, these substructures should be bluer or, at least, younger than the surrounding bulge component. However, this is not the case, as IDs exhibit very similar colors to those of their host bulges usually \citep[see][]{2004MNRAS.354..753M,2007MNRAS.379..445P}. Therefore, although gas and star formation effects must have been relevant in the formation of IDs and IRs, they may not be essential for it in many cases. This is also supported by the significant old underlying stellar component detected in most IDs and IRs \citep[][]{1998AJ....115..484B,1998AJ....116.1142B,1998MNRAS.293..343V,2004A&A...428..877K,2004MNRAS.354..753M}.

The flattened structure of IDs and IRs has been traditionally interpreted as a sign of the essential role played by gas in their formation \citep[][]{2007MNRAS.379..418C}, but our models prove that central thin rotationally-supported ICs can result from satellite disruption without supplying gas to the remnant centre. However, this is a simplified picture of the reality, as the majority of the observed IDs and IRs contain recent (or on-going) star formation, dust, and gas, a fact that clearly points to the tight relation between dissipative processes and their buildup \citep[see, \eg,][F06]{1996ApJ...471..115B,1998MNRAS.298..267V,2004MNRAS.354..753M,2005AJ....129.2636K,2007MNRAS.379..418C,2007MNRAS.379..445P}. In fact, we could expect that the inclusion of gas and star formation in the models would made the formed ICs more detectable, accounting for the star formation that minor mergers usually trigger in the centre of galaxy discs \citep{2009MNRAS.394.1713K}. So, we intend to re-run these simulations including gas and star formation processes in the near future, to determine the effects of dissipative processes in the formation of ICs.

\section{Discussion}
\label{sec:discussion}

The present models provide a novel insight into the buildup of dynamically-cold ICs through minor mergers, because the resulting IDs and IRs do not derive from primary disc material or from a gaseous component as in other studies, but from the disrupted stellar satellite material. Traditionally, minor mergers have been considered as secondary drivers in the formation of IDs or IRs in galaxies, in the sense that the rotationally-supported ICs were thought to come from the bars triggered in the discs by the encounters, not from the minor merger themselves. The accreted satellite material did not ended in these ICs except if it reached the centre undisrupted (see \S\ref{sec:introduction}). Therefore, minor mergers have been considered just as the agents inducing bars, but the bars were the processes responsible for the buildup of the ICs. 

Our models show that both processes (the induced disc resonances and the accretion of material external to the galaxy) are extremely connected, as the resulting IC is made out of satellite material and exhibits an origin related to transitory non axisymmetric distortions of the galaxy disc at the same time (\S\S\ref{sec:formation}-\ref{sec:geometrical}). The difference between our models and previous ones from the literature lies basically in the satellite characteristics, as we have found that very different orbital configurations give place to similar ICs (see Table\,\ref{tab:ics}). In our models, the discy structure and the realistic density contrast of the satellites with respect to the primary galaxies make them to be sensitive to disruption, but resistant enough to reach the remnant centre \citep[in contrast with previous models, see][]{2001A&A...367..428A,2010MNRAS.403.1009M}. This implies that, if a low-density satellite  (as a dS) were accreted by a galaxy, the existence of a prominent bulge in the primary galaxy would induce the complete disruption of the satellite and the formation of an IC out of it. However, if the accreted satellite were dense (such as a dSph or dE) or if the primary galaxy had a small bulge, the undisrupted satellite core would deposit in the centre without disrupting, increasing the spatial density in the galaxy core. 

The bar-related origin of numerous bulges and dynamically-cold ICs in galaxies is indisputable \citep[][]{2006ApJ...637..214M,2007ApJ...666..189B,2007A&A...472...63R,2009MNRAS.400.1706A,2009MNRAS.394...67A,2010MNRAS.407.1433A,2009MNRAS.395..537B,2010ApJ...716..942F}, but it is also true that more than one third of them in Sa-Sb galaxies are apparently unrelated to bars (see references in \S\ref{sec:introduction}). Considering the relevance that minor mergers seem to have had in galaxy evolution \citep[see][]{2009ApJ...697.1971J,2010A&A...518A..20L,2010ApJ...710.1170L}, it is probable that even many bars may have been induced by minor mergers in turn.  So, the present models provide a feasible explanation to the existence of old, pure stellar IDs and IRs in unbarred galaxies \citep[specially to the counter-rotating cases, see][]{2007MNRAS.376..997J,2010arXiv1004.2190M}.

The mixing of material observed in the models (due to the merger and/or to the triggered oval distortions) indicates that it would be very difficult to prove the minor merger origin of the resulting ICs by disentangling the stellar population with an external origin from the underlying stars. However, the minor merger origin of the ICs in some galaxies is obvious, mainly if the IC is counter-rotating or it is harboured by a non-relaxed host \citep[see][]{2000AJ....120..703H,2002ApJ...579..592G,2010MNRAS.402.2140S,2011arXiv1103.1692S}. 

The key role of minor mergers in the growth of bulges of spiral galaxies is evident, as relics of disrupted satellites and on-going minor mergers are frequently found in nearby galaxies, indicating that these processes are extremely common  \citep[]{2007ApJ...660.1264M,2008ApJ...689..184M,2009ApJ...692..955M,2010AJ....140..962M,2010ASPC..423..342K}. Therefore, the present models suggest that the majority of the ICs found in spiral galaxies could have had a minor-merger related origin (independently on whether the minor merger triggers a noticeable bar or not), and that the role of minor mergers in the formation of ICs may have been much more complex than just bar triggering, as traditionally assumed.

\section{Summary and conclusions}
\label{sec:conclusions}

We have investigated the capability of minor mergers to trigger the formation of IDs and IRs in spiral galaxies through collisionless N-body simulations. We have extended the simulations of minor mergers onto disc galaxies presented in EM06, sampling a wider parameter space of initial conditions. Different orbits and mass ratios have been considered, as well as two different models for the primary disc galaxy (Sab or Sc galaxy). 

All the simulated minor mergers have developed thin rotationally-supported ICs out of disrupted stellar satellite material, with scale lengths analogous to those observed in real IDs and IRs. The resulting ICs are highly aligned to the main galactic plane of the remnant, as the original non-spheroidal potential of the primary galaxy makes the satellite orbit to evolve to its privileged plane prior to disruption. This fact provides a possible explanation for the low misalignment observed in the ICs found in Sa-Sb galaxies as compared to those in E-S0's. No relevant counterparts of these dynamically-cold ICs are obtained in the remnant material originally belonging to the primary galaxy.

The geometrical analysis of these ICs reveals a wide morphological zoo of ICs, similar to the observed ones in Sa-Sc galaxies through direct imaging or through unsharp masking, such as IDs, IRs, pseudo-rings, nested IDs, spiral structure, and different combinations of them. No noticeable bars have been formed either in the primary disc or in the IC made out of accreted satellite material. The structural and kinematic properties of these ICs are analogous to those observed in real galaxies as well. The existence of these ICs cannot be derived directly from global surface density maps and profiles, but it can be deduced from the characteristic features that they imprint to the isophotal profiles and kinematic maps of the final remnants, analogously to many observational cases. 

Key points to explain the formation of these IDs and IRs in our simulations without requiring significant bars or dissipative components are the structure and density of the satellites, as well as the existence of a prominent bulge in the primary galaxy. The realistic satellite-to-primary galaxy density ratios make the satellites to be more sensible to orbital circularization and disruption than those used in previous simulations. 

Combined with the disc resonances induced by the encounter, these three processes (satellite disruption, orbital circularization, and coupling of disruption with merger-triggered resonances in the disc) give place to highly aligned rotationally-supported ICs in the remnants centre. The existence of big bulges in the primary galaxies and long-lasting decaying orbits ensure a more efficient satellite disruption, producing thinner ICs. This implies that, if a low-density satellite (such as a dS) were accreted by a galaxy with a prominent bulge, it would result in the complete disruption of the satellite and the formation of a rotationally-supported flat IC. However, if the galaxy accreted were a high-density satellite (such as a dSph or dE) or if the primary galaxy had a small bulge, the undisrupted satellite core is expected to sink to the galaxy centre without disrupting completely, contributing to the formation of a central bulge. In this sense, minor mergers could account for the existence of old, pure stellar IDs and IRs in many unbarred galaxies (specially the counter-rotating ones).

Traditionally, minor mergers have been considered just as secondary agents in the formation of dynamically-cold ICs in galaxies, only responsible for inducing the bars in the galaxy discs that finally give place to these ICs. Our models suggest that the majority of the ICs found in spiral galaxies could have had a minor-merger related origin (independently on whether the minor merger triggers a bar or not), and that the role of minor mergers in the formation of ICs may have been much more complex than just bar triggering. In conclusion, the present models prove that minor mergers are an extremely efficient mechanism to form rotationally-supported stellar ICs in spiral galaxies, neither requiring strong dissipation nor the development of strong bars. 

\begin{acknowledgements}
We thank our anonymous referee whose suggestions helped us to improve the clarity and presentation of the results. We are also very grateful to P.~Erwin, L.~S.~Sparke, and to the American Astronomical Society for the permission to reproduce some figures from \citet{2003ApJS..146..299E}. Supported by the Spanish Ministry of Science and Innovation (MICINN) under projects AYA2009-10368, AYA2006-12955, and AYA2009-11137, and by the Madrid Regional Government through the AstroMadrid Project (CAM S2009/ESP-1496, http://www.laeff.cab.inta-csic.es/projects/astromadrid/main/index.php). Funded by the Spanish MICINN under the Consolider-Ingenio 2010 Program grant CSD2006-00070: "First Science with the GTC" (http://www.iac.es/consolider-ingenio-gtc/). ACGG is a Ramon y Cajal Fellow of the Spanish MICINN. 
\end{acknowledgements}

\bibliographystyle{aa}{}
\bibliography{elic0709_def.bib}{}

\end{document}